\def\smallskip{\vskip 8pt}
\def\littleskip{\vskip 6pt}

\documentstyle[psfig]{mn}
\title{Galaxy Formation and Evolution.~I. \\
       The Padua TREE-SPH code (PD-SPH) }
\author[Giovanni Carraro, Cesario Lia, and Cesare Chiosi]
       {Giovanni Carraro$^{1,2}$, Cesario Lia$^{2}$, and Cesare Chiosi$^{1}$\\
       $^{1}$ Dipartimento di Astronomia, Universit\`a di Padova, Vicolo
dell'Osservatorio 5, I-35022 Padova, Italy\\
       $^{2}$ SISSA-ISAS, via Beirut 2-4, I-34013 Trieste, Italy  }
\date{Accepted.......
      Received.......;
      in original form......}
\pubyear{1996}
\begin{document}
\maketitle
\title{The PD-SPH code}


\begin{abstract}
In this paper we report on PD-SPH the new tree-sph code developed in
Padua.
The main features of the code are described and the results of a new 
and independent series of  1-D and 3-D tests are shown. 
The paper is mainly dedicated to the presentation of the code and to 
the critical discussion of its performances.
In particular great attention is devoted
to the convergency analysis. 
The code is highly adaptive in space and time by means of
individual smoothing lengths and individual time steps.
At present it contains both 
dark and baryonic matter, this latter in form of gas and stars, 
cooling, thermal conduction, star formation, and 
feed-back from Type I and II supernovae, stellar winds,
and ultraviolet flux from massive stars, and finally chemical enrichment.
New cooling rates that  depend on the metal abundance of
the interstellar medium are employed, and the differences with
respect to the standard ones are outlined.
Finally, we show the simulation of the dynamical and chemical
evolution of a disk-like galaxy with and without feed-back.
The code is suitably designed  to study in a global fashion the problem
of formation and evolution of elliptical galaxies, and in particular to feed
a spectro--photometric code  from which the integrated
spectra, magnitudes, and colors (together with their spatial gradients) can be
derived. 
\end{abstract}

\begin{keywords}
SPH, Hydrodynamics, Galaxy formation and evolution
\end{keywords}

\section{Introduction}

The conventional picture of star formation in elliptical galaxies is one in 
which the galaxies and their stellar content formed early in the universe and
have evolved quiescently ever since. This view is supported by the apparent
uniformity of elliptical galaxies in photometric and chemical appearance 
(cf. Matteucci 1997 for a recent review) and the existence of scaling 
relations, e.g. the fundamental plane (cf. Bender 1997).  In contrast, 
the close scrutiny of nearby elliptical galaxies makes 
evident a large variety of
morphological and kinematic peculiarities and occurrence of star formation 
in a recent  past (Schweizer et al. 1990, Schweizer \& Seitzer 1992, 
Rampazzo et al. 1996). All this leads to a different picture in which 
elliptical galaxies are formed by mergers and/or accretion of smaller units 
over a time scale  comparable to the Hubble time.   
Furthermore, strong evolution in the population of early type galaxies
has been reported by Kauffmann et al. (1996) which has been considered to
support the hierarchical galaxy formation model (Kauffmann
et al. 1993, Baugh et al. 1996).

Tracing back the formation mechanism of elliptical galaxies from the bulk 
of their chemo-spectro-photometric properties is a cumbersome affair 
because  studies of stellar populations in integrated light reveal only 
luminosity weighted ages and metallicities, and are
ultimately unable to distinguish between episodic (perhaps recurrent) 
and monolithic histories of star formation and between star formation 
histories in isolation or  interaction. 

 As nowadays most of the properties of elliptical galaxies
have been studied with sophisticated chemo-spectro-photometric models 
in which the dynamical process of galaxy formation is reduced to assuming 
either the closed box or infall scheme and a suitable law of star formation
(Arimoto \& Yoshii 1987, 1989; 
Bruzual \& Charlot 1993; Bressan et al. 1994; Worthey 1994; 
Einsel et al. 1995; Tantalo et al. 1996; Bressan et al. 1996; 
Gibson 1996a,b,c; Gibson 1997; Gibson \& Matteucci 1996; Tantalo et al. 1997).

In contrast, the highly sophisticated dynamical models of galaxy formation 
(Hernquist \& Katz 1989; Davis et al. 1992; Katz 1992; 
Steinmetz \& M\"uller 1993; 
Navarro \& White 1993; Nelson \& Papaloizou 1994; Katz, Weinberg \&
Hernquist 1995; Navarro \& Steinmetz 1997; Haehnelt et al. 1996a,b; 
 Steinmetz \& Mueller 1995; Navarro et al. 1996; Steinmetz 
1996a,b and references), owing to the complexity of the problem are still
somewhat unable to make  detailed predictions about the 
chemo-spectro-photometric properties of the galaxies in question. 

Attempts  to bridge
the two aspects of the same problem are by Theis et al. (1992 and references)
and most recently by Contardo, Steinmetz \& Fritze-v-Alvensleben (1997 in
preparation).

This paper is the first of a series dedicated to the study of this complex 
problem in a self-consistent fashion, in which the formation and evolution
of galaxies, together with their chemo-spectro-photometric properties,
stem from a unique model able to predict a number of observable quantities to be
compared with real data. The project aims to build dynamical models of
disk-like and elliptical galaxies by means of the SPH (Smoothed Particle 
Hydrodynamics) technique, in which the effects of different initial conditions
and basic physical processes, such as star formation,  heating
  of the gas by various mechanisms (supernova explosions, stellar winds,
UV fluxes),  cooling of this by radiative atomic and molecular agents,
interplay between dynamics and thermodynamics (feed-back), and 
chemical enrichment of the inter-stellar medium are taken into 
account. The models allow for the presence of dark matter and the effect
of this on the gravitational field, which is described using Barnes \&
Hut (1986) treecode. The rates of star formation and 
chemical enrichment as function of space and time resulting from the 
dynamical models are meant to feed chemical-spectro-photometric
models which should provide the kind of data (luminosities, integrated
spectral energy distributions, broad band colors, line strength indices, etc.)
to be compared with 
real observations for nearby and high redshift galaxies.

In this paper we present the first step toward this articulated analysis of 
the problem, i.e. the tool to construct dynamical models, and 
describe in  some
detail the numerical algorithm and the key physical ingredients.
The dynamical models are based on the 
SPH technique, one of the most popular and efficient tool of modern
studies of numerical hydrodynamics, and in astrophysical context of
structures  and galaxy formation. In the following we will refer to 
the SPH code we have developed as the Padua SPH (PD-SPH).

When SPH is coupled with an efficient scheme to compute gravity
(like tree codes), it becomes
a  powerful tool to investigate physical situations of enormous complexity
such as for instance, the formation and evolution of a galaxy 
(in the isolation picture) and/or the
interaction-merge of galaxies (in the hierarchical scenario).  
 The lagrangian nature of the method allows us 
to follow  the evolution of physical quantities at all scales, provided that
sufficient resolution is ensured.

The plan of the paper is as follows. 
Section~2 gives a detailed description of the code. 
Section~3 discusses
tests in one dimension, i.e. the results for the Riemann problem and the
 convergency analysis, whereas section~4 contains 
tests performed in three dimensions. Section~5 
presents the first astrophysical application, i.e. the collapse of a
galaxy made of gas and dark matter in presence of initial solid body rotation.
Section~6 thoroughly presents non--adiabatic processes, such as thermal
conduction, radiative cooling, and heating by Type I and II 
supernova explosions, and 
stellar winds and ultraviolet radiation from massive stars that have used as 
source or sink of energy in our models.
We adopt cooling rates that take the effect of metallicity into account. 
In Section 7 we 
describe the results obtained from our cooling prescription and
compare them with those from metal independent cooling rates.
Section~8 describes our prescriptions for the star formation rate and 
feed-back (rate of energy input from supernovae, stellar wind, and UV
emission), and rate of chemical enrichment. Section 9 presents the results
for two disk-like galaxies: the first case is with no feed-back of any type,
whereas the second case is with  feed-back. These models are not meant
to represent real galaxies, first because the initial conditions 
are not derived from a cosmological context, second because the 
total number of particles is small owing to limitations in the computing
facilities. The models are indeed meant to test the code response to
various physical inputs. Nevertheless, despite the present limitations,
the results are very promising. Finally, Section 10 
presents some concluding remarks and outline the perspectives of future
studies.

\section{The PD-SPH code}
PD-SPH is a TREE--SPH code, written in Fortran 90 
(cf. Carraro 1995 \& Lia 1996), conceptually  similar to those described by 
Hernquist \& Katz (1989) and Nelson \& Papaloizou (1994).
A 1-D version of the SPH code is in parallel form, and it 
has already been
tested on the Cray T3D parallel computer hosted by CINECA
(Beninc\`a \& Carraro 1995).
 
Schematically,  PD-SPH
uses $\rm SPH$ to solve the equations of motion for the gas component, and the
Barnes \& Hut (1986) octo-tree to compute the gravitational interactions.
In this section we  describe in some details the structure of 
the code and its basic ingredients.

It is widely known that SPH is a method standing on two basic ingredients:
an interpolation using a kernel,  and a Monte--Carlo evaluation of the 
physical quantities (Monaghan 1985). 
All relevant variables are interpolated in the following way:

\begin{equation}
\left\langle f({\vec r})\right\rangle = \int W({\vec r}-{\vec r}',h)
f({\vec r}')\,d{\vec r}',
\end{equation}

\noindent
where the integral is extended over the whole space and $W$ is a 
function generally referred to as the interpolating kernel;
$h$ is the {\em smoothing length} determining the spatial 
region within which 
variables are smoothed,  and governing 
the spatial resolution of the method.

The kernel is a spherically symmetric function strongly peaked 
at $\vec r = \vec r^{'}$. 
It is easy to prove (Benz 1990) that these properties make SPH a 
second order technique, in which the estimate of any function is   
given by

\begin{equation}
\left\langle f({\vec r})\right\rangle=f({\vec r})+c( 
      \nabla ^2 f)h^2+ O(h^3).
\end{equation}

\noindent
This estimate is made  in  lagrangian formalism summing up the 
contributions of all the
particles within $2 \times h$, which represent the so--called neighbors.  

\subsection{Interpolating kernel}
Three types of kernel can basically be found  in literature: exponential,
gaussian and spline; see Benz (1990) for more details. We have adopted 
 the spline--kernel of Monaghan \& Lattanzio, (1985):

\begin{equation}
W(r,h) = \frac{1}{\pi h^{3}} \left\{ \begin{array}{ll}
1 - \frac{3}{2} u^{2} + \frac{3}{4} u^{3}, & \mbox{se $0 \leq u \leq 1$;}\\
\frac{1}{4} (2 - u)^{3},                   & \mbox{se $1 \leq u \leq 2$;}\\
0.					   & \mbox{otherwise,}  
		                     \end{array}
		             \right.
\label{kspl}
\end{equation}

\noindent
and the supergaussian--kernel of Gingold \& Monaghan, (1982):

\begin{equation}
W(r,h) =  \frac{1}{\pi^{3/2} h^{3}} (\frac{5}{2} - u)
 exp(-u^{2}),
\label{supgaus}
\end{equation}
where $u=|{\vec r}|/h$.

The super--gaussian kernel is built up forcing the second order term in 
eq. (2) to be zero, It can be shown that this kernel corresponds to
 a fourth order interpolation. This kernel becomes negative 
for $r/h > \sqrt{5/2}$ (Benz 1990) and does not find an immediate 
 physical interpretation. 

Moreover,  while the spline kernel is defined on a compact support, the 
super--gaussian kernel 
needs an artificial cut--off to limit the number of neighbors utilized in
the estimate of the physical quantities. 
First, we checked both kernels (see Section~3 for details) and finally
adopted the spline-kernel. In our code it is stored as a look--up table.

\subsection{Evolution of the smoothing length and the search of neighbor
particles }

In SPH the spatial resolution is fixed by the smoothing
length $h$. 
In our code $h$ is kept variable both in space and 
time according to Benz (1990)

\begin{equation}
\frac{dh}{dt}=-\frac{1}{3} \frac{h_{i}}{\rho_{i}} \frac{d\rho_{i}}{dt}
=\frac{1}{3}h {\vec \nabla}  {\vec v}.
\end{equation}

\noindent
This  equation is added to the set of equations to be solved at any
time--step.
The main drawback of this approach is that in situations in which strong
density gradients are present, it is not possible to keep under control
the number of neighbors. Typically this number should be about 40--50
(Steinmetz \& M\"uller 1993).

One possible solution is to artificially increase or decrease $h$ 
in order to keep fixed the neighbor number, but this involves
several iterations if -- as in our case --  the tree-code
is used to look for neighbors. As a consequence of this, 
the computational time
gets often unacceptably long. 

To cope with this difficulty, we have decided to maintain the above
 scheme when the neighbor number is lower 
than 40, and   switch to the Nelson \& Papaloizou (1994) formalism
when the neighbor number is
greater than 80. Accordingly the new smoothing length $h$ is

\begin{equation}
h_{i} = \frac{1}{N_{far}} \sum_{n=1}^{N_{far}} \frac{1}{2} | \vec r _{i}
- \vec r_{n} |
\end{equation}

\noindent
where  $N_{far}$ are the $n$  most distant nearest neighbors. We
assume $N_{far} = 10$.

There is another drawback in the Benz (1990) formalism that deserves some 
attention. The space--time
variation of the smoothing length generally hampers the strict conservation of
energy when terms involving the space derivatives of $h$ are not included.
$\nabla h$ terms are essentially small correction 
(cf. the thorough discussion by Nelson \& Papaloizou 1994), which do
not significantly improve upon the conservation of energy.

Moreover the momentum conservation has been secured at an acceptable level of confidence
also by symmetrizing the search of the nearest neighbors.
In brief, a candidate particle
$i$ is considered as a neighbor of $j$ if the following conditions are met

\[
|{\vec r}_i-{\vec r}_j|< 2 \cdot h_i 
\]

\noindent
or

\[
|{\vec r}_i-{\vec r}_j|< 2 \cdot h_j .
\]

\begin{figure*}
\centerline{\psfig{file=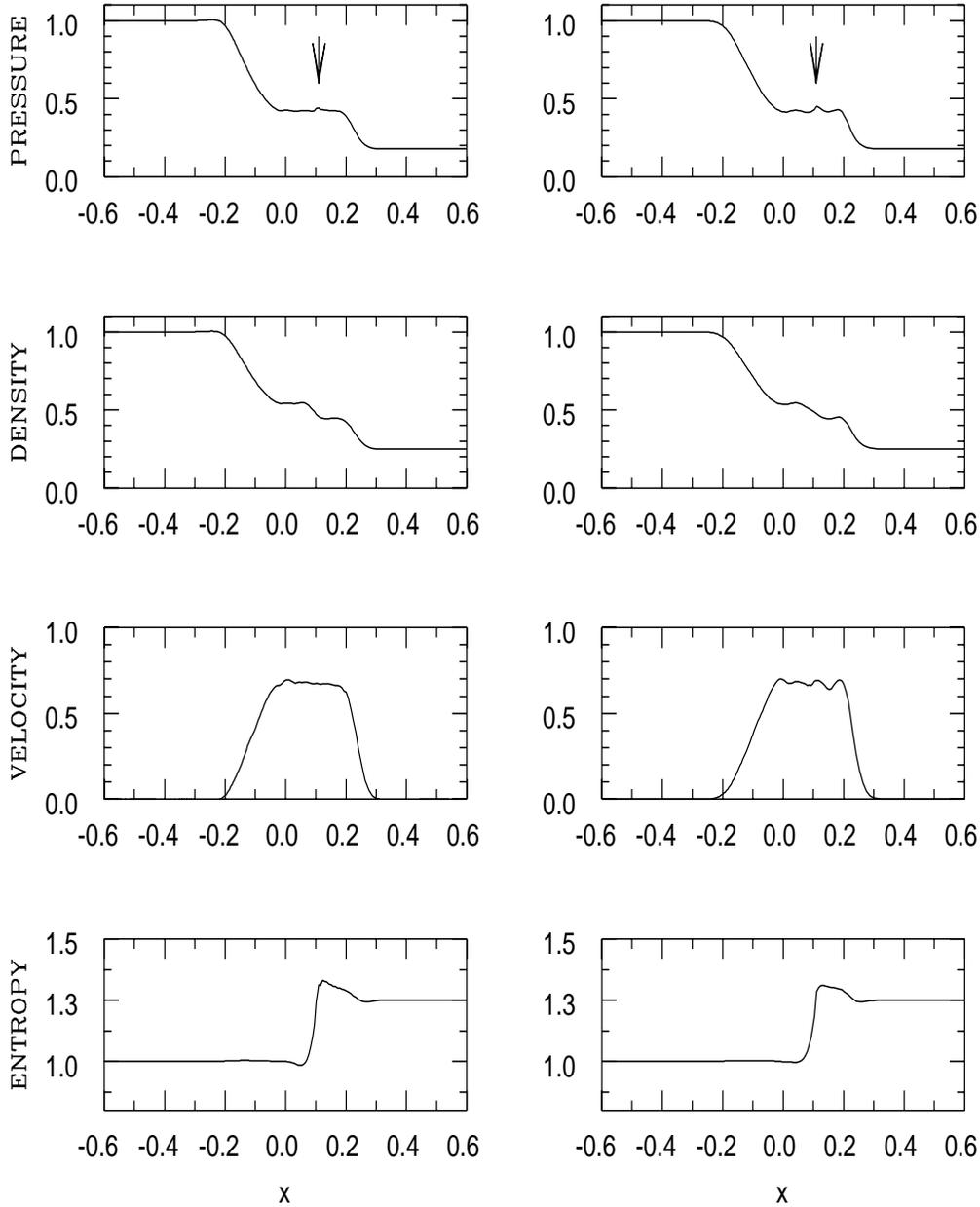,height=18cm,width=14cm}}
\caption{Result of the shock tube problem using super-gaussian kernel 
(left panel) and spline kernel (right panel). In both cases 400 particles
have been used. The arrow indicates the so--called blip in the pressure
profile}
\label{tube}
\end{figure*}

\subsection{Hydrodynamical equations}

In PD-SPH, the  particle forces and the specific energy are computed by
means of the following equations

\[
\frac{d \vec v_i}{d t} = - \sum_{j=1}^N m_j 
\left (\frac{\sqrt{P_i P_j}}{\rho_i \rho_j }  + \Pi_{ij}\right ) \times
\]
\begin{equation}
\frac{1}{2} 
\left ( {\vec \nabla }_i W(r_{ij},h_i)+ \vec \nabla_i W(r_{ij},h_j)\right ))
\label{eq_velocity}
\end{equation}

\noindent
and

\[
\frac{du_i}{dt}=\sum_{j=1}^N m_j \left 
(\frac{\sqrt{P_{i}P_{j}}}{\rho_{i}\rho_{j}}
 + \frac{1}{2} \Pi_{ij}\right ){\bf v}_{ij} \times
\]
\begin{equation}
\frac{1}{2} 
\left ( {\vec \nabla}_{i} W(r_{ij},h_i)+ \vec \nabla_i 
W(r_{ij},h_j)\right ) + \Gamma_R -\frac {\Lambda_C}{\rho},
\label{eq_energy}
\end{equation}

\noindent
in which we adopt the arithmetic mean for the pressure gradient
as in Hernquist \& Katz (1989). Furthermore, in eq. (\ref{eq_energy})
the first term 
represents the heating rate of mechanical nature (it is shortly indicated
as $\Gamma_M$),  
the second term  $\Gamma_R$ is  the total heating rate
from all sources apart from the mechanical ones, and the third term
 $\Lambda_C / \rho $ is the total
cooling rate by many physical agents. These latter
two will  be discussed below in more detail.
 In the above equations $\vec v_{ij} = \vec v_{i} - \vec v_{j}$, and
$\Pi_{ij}$ is the viscosity tensor defined as

\begin{equation}
\Pi_{ij} = \left\{ \begin{array}{ll}
\frac{-\alpha c_{ij} \mu_{ij} + \beta \mu_{ij}^{2}}{\rho_{ij}}, &
\mbox{if $(\vec v_{i} - \vec v_{j}) \cdot (\vec r_{i} - \vec r_{j})  > 
0$} \\
0.                                                              &
\mbox{otherwise,}
                 \end{array}
           \right.
\end{equation}

\noindent
where

\begin{equation}
\mu_{ij}=\frac{h_{ij}({\vec v}_i-{\vec v}_j)({\vec r}_i-{\vec r}_j)}{|{\vec 
r}_i-{\vec r}_j|^2 + \epsilon h_{ij}^2}  .
\end{equation}

\noindent
Here $c_{ij}=0.5(c_i+c_j)$ is the sound speed, 
$h_{ij}=0.5(h_i+h_j)$, and $\alpha$ and $\beta$
are the viscosity parameters, usually set to 1.0 and 2.0, respectively.
The factor 
$\epsilon$ is fixed to 0.01 and is meant to avoid divergencies.

As amply discussed by Navarro \& Steinmetz (1997) this formulation
has the disadvantage of not vanishing in the case of shear dominated
flows, when $\vec \nabla \cdot \vec v = 0$ and $\vec \nabla \times
\vec v \neq 0$. In such a  case, spurious shear viscosity can
develop, mainly in simulations involving a small number of particles.
To reduce the shear component we adopt the Balsara (1995)
formulation of the viscosity tensor 

\begin{equation}
\tilde \Pi = \Pi_{ij} \times \frac{f_{i} + f_{j}}{2},
\end{equation}
 
\noindent
where $f_i$ is a suitable function defined as

\begin{equation}
f_i=\frac{|<{\vec \nabla} \cdot  {\vec v}>_i|}{|<{\vec \nabla} \cdot {\vec 
v}>_i| + |<{\vec \nabla }\times {\vec v}>_i|+\eta c_i/h_i},
\end{equation}

\noindent
and $\eta \approx 10^{-4}$ is a parameter meant to prevent  
numerical divergencies.

\subsection{Time integration and stability criteria}
Particle positions and velocities are updated, as in 
Hernquist \& Katz (1989), using the leapfrog 
integrator and the multiple time--step scheme.
The integration is of the second order in time, and proceeds
as in the classical scheme.
 
Firstly an estimate of the velocity $\tilde {\vec v}_i^{n+1/2}$ is obtained
from 

\begin{equation}
\tilde {\vec v}_i^{n+1/2}= {\vec v}_i^n +0.5 \Delta t_{i} {\vec a}_i^{n-1/2}.
\end{equation} 

\noindent
This is used to compute time-centered accelerations, ${\vec 
a}_i^{n+1/2}$, from which particle velocities and positions  are 
updated

\begin{equation}
{\vec v}_i^{n+1}={\vec v}_i^n +\Delta t_{i} {\vec a}_i^{n+1/2} 
\end{equation}

\begin{equation}
{\vec r}_i^{n+1/2}={\vec r}_i^{n-1/2}+\Delta t_{i} {\vec v}_i^n   . 
\end{equation}

\noindent
The energy equation is explicitly solved, unless sink or source terms 
are present, and particle energies are advanced in the same manner
as positions.
  
Time steps are calculated according to the Courant condition

\begin{equation} 
\Delta t_{C,i} = {\cal C} \frac{h_{i}}{h_{i} \left | \vec \nabla \cdot \vec 
v_{i}  
\right | + c_{i} + 1.2(\alpha c_{i} + \beta max_{j} \left | \mu_{ij} 
\right |)}   ,
\end{equation}

\noindent 
with $\cal C$ $\approx 0.3$.

In presence of  gravity, a more stringent condition on  the time steps is
required. According to Katz, Weinberg \& Hernquist (1995), the additional
criterion has to be satisfied

\begin{equation}
\Delta t_{G,i} = \eta \cdot MIN (\frac {\eta \epsilon}{|\vec v |},
(\frac{\epsilon}{|\vec a|})^{1/2}),
\end{equation}

\noindent
where $\epsilon$ is the gravitational softening parameter and
 $\eta$ is another parameter usually set to 0.5. 
The final time step
to be adopted is the smallest of the two

\begin{equation}
\Delta t_{i} = MIN (\Delta t_{C,i},\Delta t_{G,i}).
\end{equation}

\noindent
While the use of the multiple time step scheme may significantly 
speed up   the code (Steinmetz 1995), sometimes 
the interpolations of non active particles may affect the
energy conservation (see Section~4).

\begin{figure}
\centerline{\psfig{file=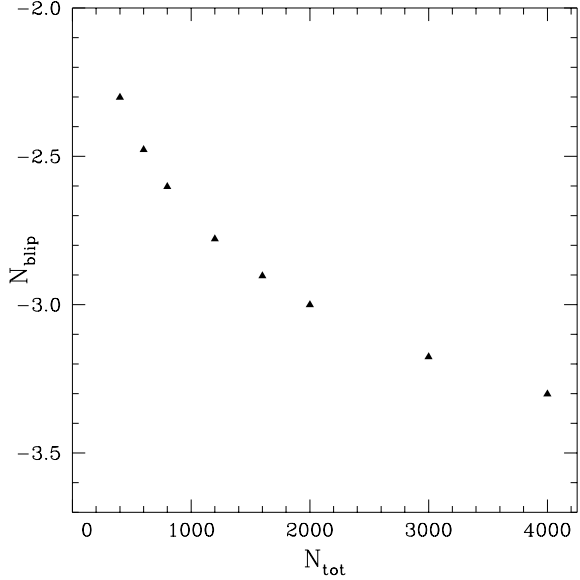,height=8cm,width=8cm}}
\caption{The number of particles involved in the {\it blip} versus the 
total number of particles in the simulation. $N_{blip}$ is on 
logarithmic scale}
\label{blip}
\end{figure}

\begin{figure}
\centerline{\psfig{file=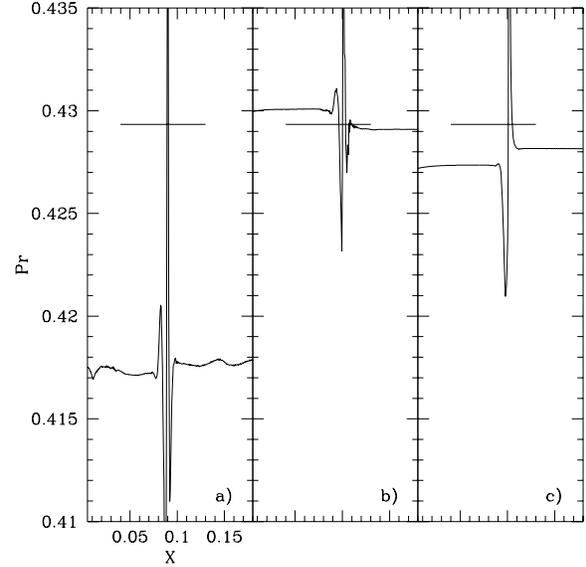,height=8cm,width=8cm}}
\caption{Blow-up  of the contact discontinuity in the pressure profile. 
The left panel (a)
shows
the results obtained from adopting the super--gaussian kernel and 
the  force term from
Gingold \& Monaghan (1983). The central (b) and right (c) panels show the 
same but for standard formalism for the pressure gradient
 and the super--gaussian and
spline kernel, respectively. The dashed line is the analytical solution }
\label{blow}
\end{figure}

\begin{figure}
\centerline{\psfig{file=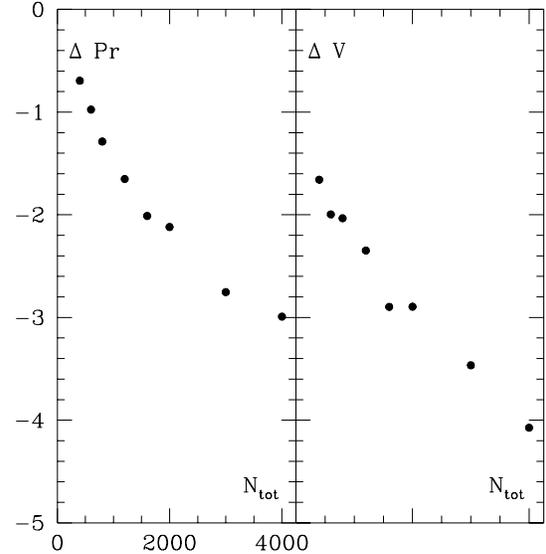,height=8cm,width=8cm}}
\caption{Convergency test. The differences $\Delta$ between the numerical
and analytical solutions for pressure and velocity profiles as a function of
the total number of particle in the simulation}
\label{conv_1}
\end{figure}

\section{1-D numerical tests }

The classical 1-D test to which SPH codes are compared is the Riemann 
shock tube problem. 
We use this test not only to check whether our code works properly, but 
also to analyse its convergency performances.
The classical reference for this test is Gingold \& Monaghan (1983),
to whom we  refer for the initial conditions of the problem.
The time--steps and the smoothing lengths are constant, and fixed to 0.05
and 0.025, respectively. In other words, the smoothing lengths are set to be
two times the maximum inter--particles separation.
The particle masses (0.003125 in our case)
are obviously chosen in such a  way that the density and pressure profiles 
are matched 
(Gingold \& Monaghan, 1983). The kernel is suitably normalized using the
 normalization constant  $\frac{2}{3h}$  
(see for instance Hernquist \& Katz 1989.) 

\begin{figure*}
\centerline{\psfig{file=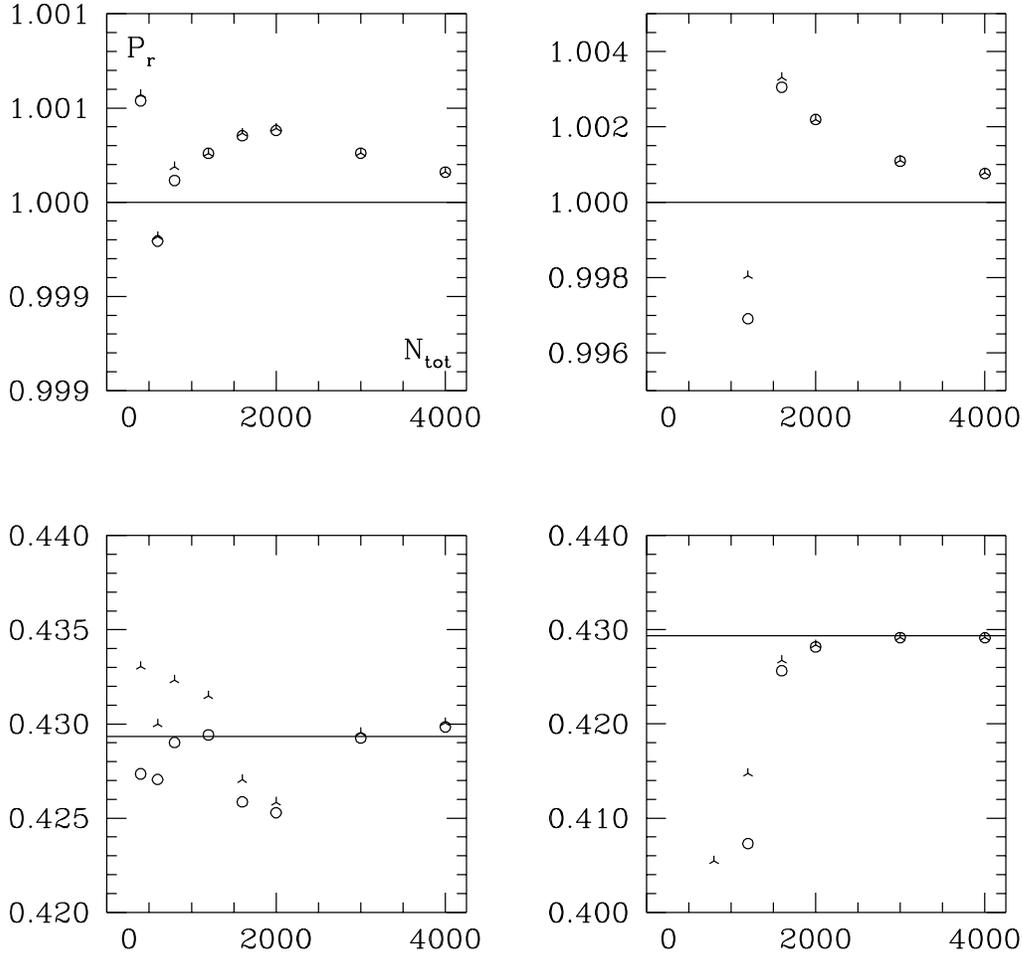,height=14cm,width=14cm}}
\caption{Convergency test. The differences $\Delta$ between the numerical
and analytical solution in each of the four regions defined in the text.
Open circles and crosses are for the minimum and maximum pressure. 
The continuous line represents the analytical solution}
\label{conv_2}
\end{figure*}

In Fig.~\ref{tube} we show the results for pressure, density, 
velocity,  and entropy
of a simulation of the shock tube problem obtained using  400 particles 
and both
the spline and  super-gaussian kernel (right and left panels, respectively).
The aim is to 
check whether using different kernels it might possible to eliminate the so
called {\it blip} in the pressure profile (indicated 
by an arrow in Fig.~\ref{tube}).
The nature of the {\it blip} is well known, and it is the signature of
the contact discontinuity in the shock. In practice the pressure is 
discontinuous at this point, and the correspondent smoothed value
$<Pr>$ is expected to be somewhat lower and higher at the left and right side 
of the contact discontinuity, respectively (Gingold \& Monaghan 1983).
Even if in our simulation  the shock is broadened over a $4 \times h$ region, 
the {\it blip} occurs as the signature of the 
unphysical step in the pressure.

We find  that even  using the super-gaussian kernel the {\it blip} albeit 
smoothed out does not disappear. In contrast to Gingold
and Monaghan (1983), we are not able to eliminate the {\it blip} even
using their new formulation for the pressure force 

\begin{equation}
\frac{\nabla Pr}{\rho} = \rho^{\gamma -2} \left [ \nabla (Pr \rho) + 
(\gamma - 1) Pr \nabla \rho \right ].
\end{equation}

\noindent
Instead of trying other expressions for the pressure gradient, once adopted
 a certain combination of kernel and pressure gradient, we
prefer to run simulations with a different number of particles to 
check whether
the SPH code can eventually recover the analytical solutions
when larger and larger numbers of particles are used.

In Fig.~\ref{blip} we show the number, $N_{blip}$, of particles 
involved in the {\it blip} as a function of the total number of particles 
$N_{tot}$ in the simulation. $N_{\it blip}$ is  the number of
particles counted in a searching 
sphere of radius equal to $2\times h$ near the {\it blip} location, and it is
normalized to the total number $N_{tot}$.
The results are as expected, in the sense that $N_{\it blip}$
 strongly decreases at increasing $N_{tot}$, or equivalently
at decreasing smoothing scale. These experiments show that the {\it blip}
survives because of the still insufficient resolution. 

In Fig.~\ref{blow} we show the results of simulations performed using different 
combinations of kernels and gradient pressure terms. The region of interest
is blown up for the sake of better understanding.
Panel (a) is for super--gaussian
kernel and the functional expression for  $\frac{\nabla P}{\rho}$ 
by Gingold \& Monaghan (1983).  Panels (b) and (c) show the same 
but using the standard
formalism for the pressure gradient, and adopting super--gaussian and 
spline kernel, respectively. In all the three panels the {\it blip} is
clearly visible. 

\begin{figure*}
\centerline{\psfig{file=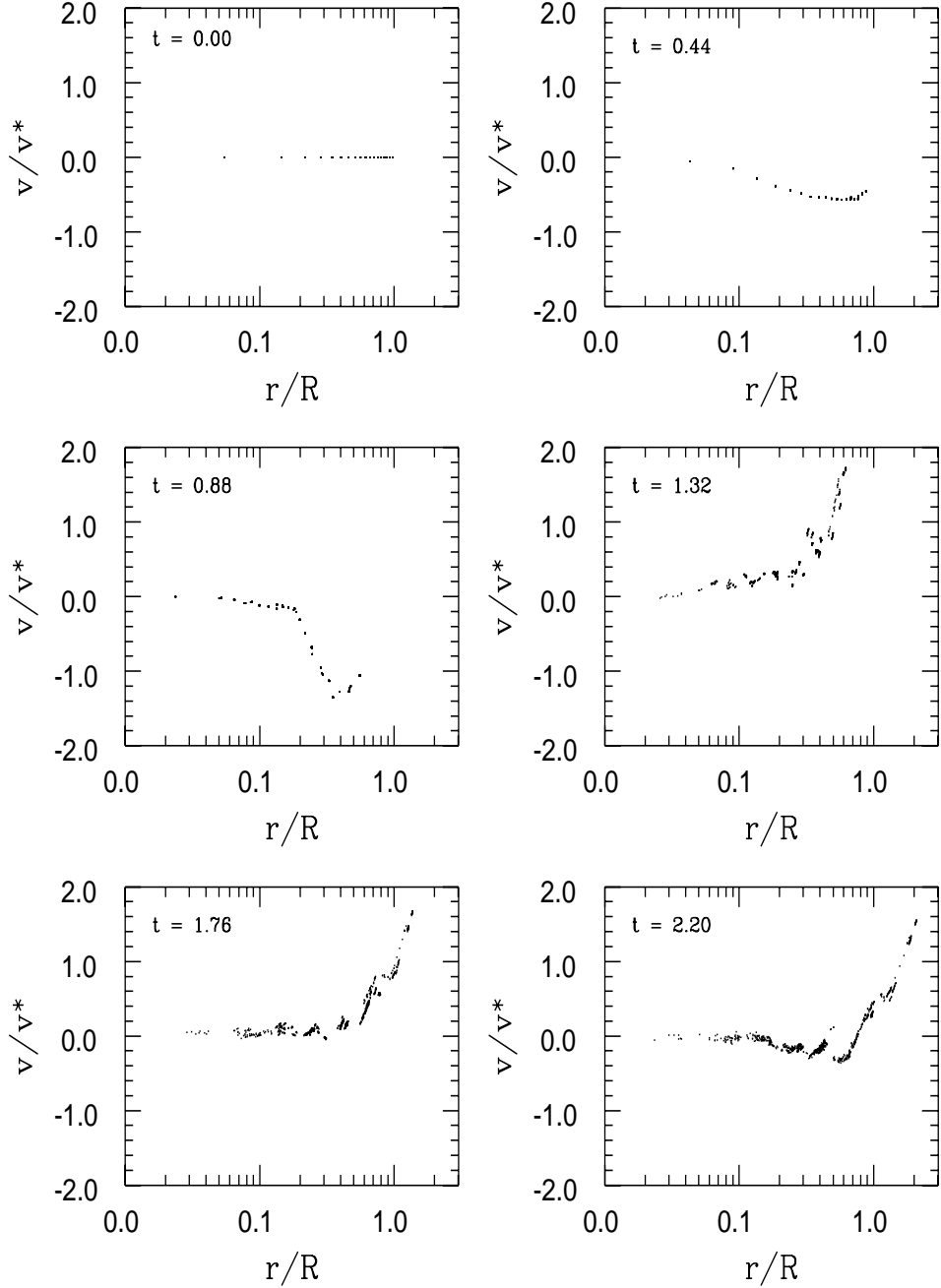,height=18cm,width=15cm}}
\caption{Adiabatic collapse: the evolution of the radial velocity 
profile (in units of
$v_{\star} = (\frac{G M}{R^{2}})^{1/2}$)  as a function of
the age. The selected ages are the same as in Hernquist \& Katz (1989) }
\label{evol_vr_ad}
\end{figure*}

\begin{figure*}
\centerline{\psfig{file=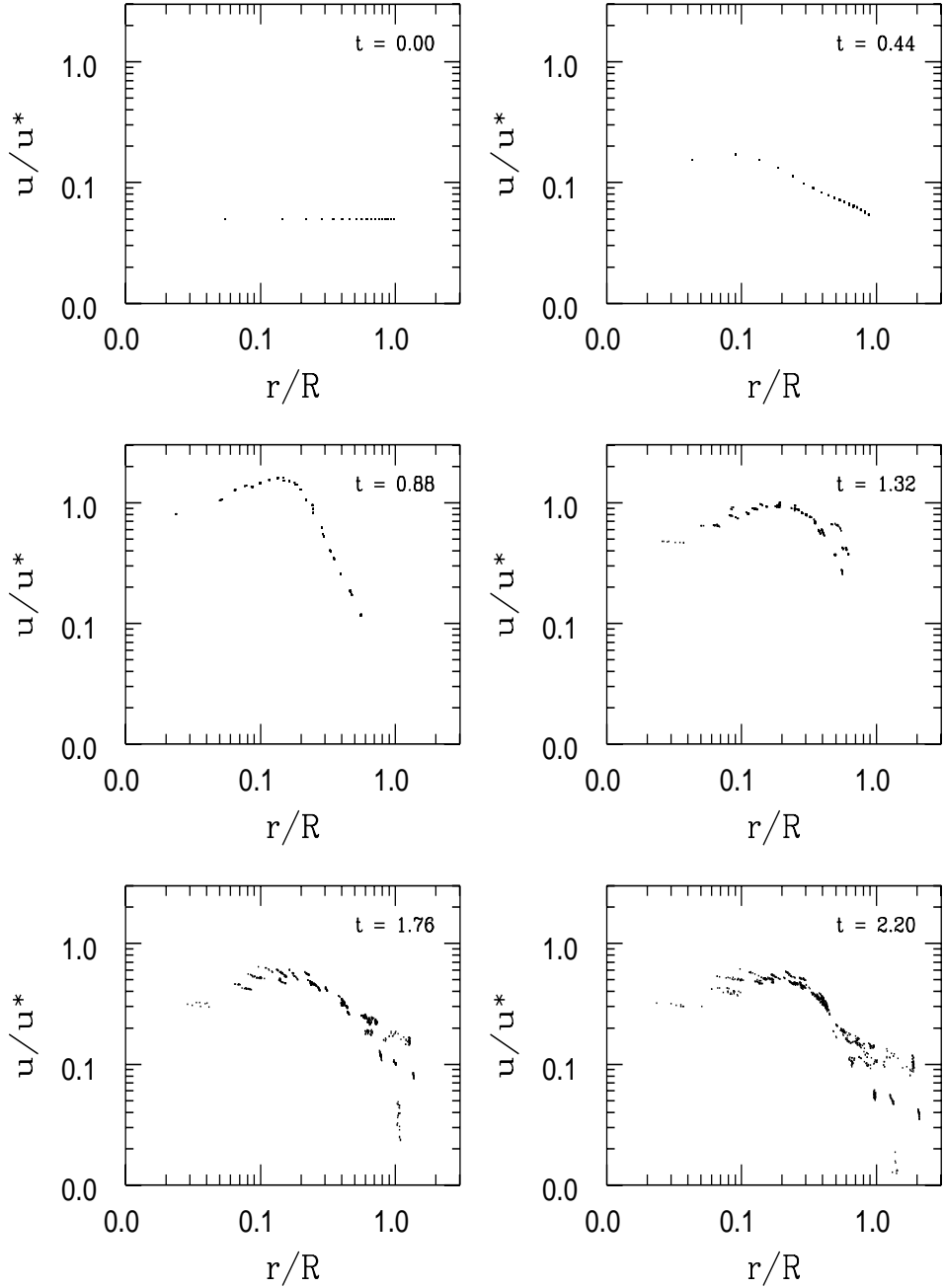,height=18cm,width=15cm}}
\caption{Adiabatic collapse: the evolution of the specific internal energy 
profile (in units of  $u_{\star} = \frac{G M}{R}$) as a function of the age. 
The selected ages are the same as in Hernquist \& Katz (1989)  }
\label{evol_ene_ad}
\end{figure*}

We notice, in particular,  that the first combination of kernel and pressure
gradient yields the best results, even if the numerical solution 
strongly departs from the analytical one. In the case with super--gaussian
kernel and standard pressure gradient (panel b), the step in 
pressure is exactly the opposite of what is expected. 
Finally, in contrast with Gingold \& Monaghan (1983),
the inclusion  of  thermal conduction (cf. Lia 1996) does not eliminate 
the pressure step.
Most likely, the {\it blip} is a feature intrinsic to the SPH
smoothing itself.

As far as testing the  convergency is concerned, we proceed as follows.
We select four regions in the pressure profile, derive 
the maximum and minimum  values from the numerical solutions
at increasing
number of particles in the simulation, and
 compare them  with
the analytical one. We choose the  pressure as test physical quantity
because of its sensitivity to the numerical technique. The pressure in fact 
is  computed from the smoothed density
and energy, instead of being directly smoothed out.

The four selected regions are the
undisturbed warm fluid at the left of the shock (1), two regions near the shock,
to the left (2) and to the right (3), respectively, 
and a region just to the right of the {\it blip} (4).  These regions 
are centered at the linear coordinate $X$ = -0.6, -0.2, +0.04 and +0.2, 
respectively.
The particles considered in the pressure evaluation are those contained
in the $2 \times h$ searching sphere.

The differences between numerical and analytical
solutions are shown in Figs.~4 and 5.

In Fig.~\ref{conv_1} we show  the pressure (left) and velocity (right) 
differences $\Delta$ between numerical
and analytical solutions at increasing
number of particles in the simulation. The plotted differences are the mean
of their values in the four test regions. As expected, the largest differences
occur in the pressure profile.
The four panels of Fig.~\ref{conv_2} show
 the maximum and minimum values of the
pressure in the four regions  at increasing total number of 
particles as indicated. In each panel the horizontal line
 shows the analytical case by definition.

All these numerical  tests prove the quick convergency of
the numerical solution to the analytical one, and compared with the similar 
experiments by other authors, show that satisfactory agreement is achieved.

Finally, we like to report that the strong double--shocks tube experiment 
(cf. Steinmetz \&
M\"uller 1993) has also been successfully performed (Carraro 1995).

\begin{figure*} 
\centerline{\psfig{file=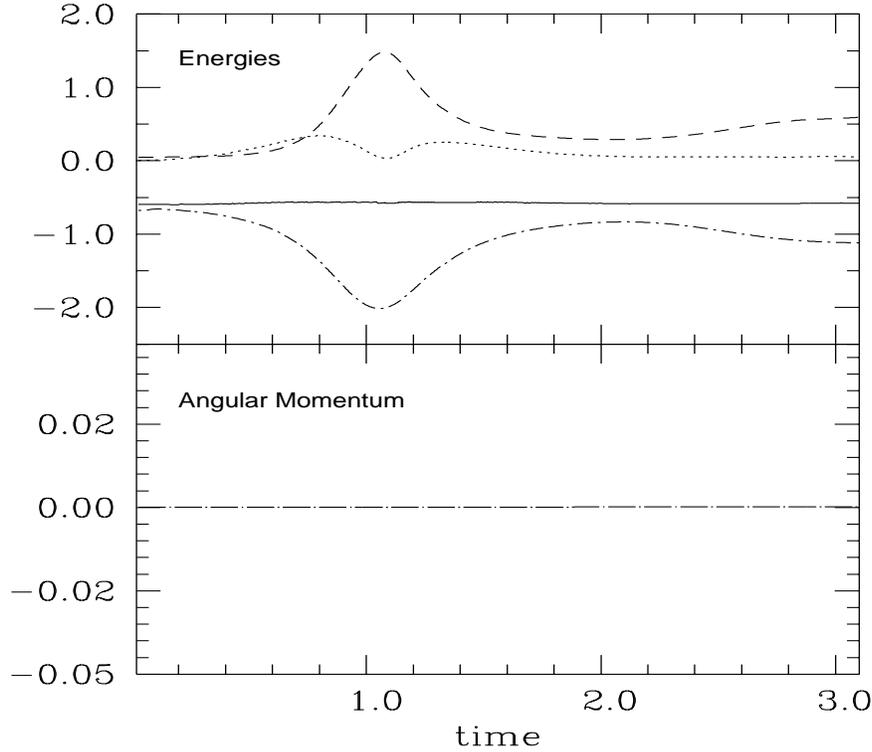,height=12cm,width=15cm}}
\caption{Adiabatic collapse.  Upper panel: conservation of the potential (dashed-dotted line), 
kinetic (dotted line),  thermal (dashed line), and 
total energy (continuous line). Lower panel: conservation of the
total angular  momentum}
\label{ene_mom_con} 
\end{figure*}

\section{ 3-D numerical tests }
In three dimensions we consider the adiabatic collapse of an 
initially non-rotating
isothermal gas sphere. This is a standard test for  SPH codes
(Hernquist \& Katz 1989;   
Steinmetz \& M\"uller 1993; Nelson \& Papaloizou 1994). 
To facilitate the  comparison of  our results with those by the
above authors, we adopt the same
initial model and the same units ($M=R=G=1$).
The system consists of a $\gamma = 5/3$ gas sphere, with an 
initially isothermal density profile:

\begin{equation}
\rho(r) = \frac{M(R)}{2\pi R^{2}} \frac{1}{r}  ,
\end{equation}

\noindent
where M(R) is the total mass inside the sphere of radius R.
Following Evrard (1988), the profile is obtained stretching an 
initially regular cubic grid by means of the
radial transformation 

\begin{equation}
r_{i}^{old} \Rightarrow r_{i}^{new} =
(\frac{r_{i}^{old}}{R})^{3}R.
\end{equation}

Alternatively it is possible to use the 
acceptance-rejection procedure by Hernquist \& Katz (1989).
The  total number of particles used in this simulation
is 2176. All the particles have the same mass.
The initial specific internal energy is set to $u = 0.05GM/R$.

\subsection{Gravity}
In PD-SPH, the newtonian forces are calculated  using the Barnes--Hut
tree-code (Barnes \& Hut 1986). Tree-code is also used to perform
the search of neighbors. We always include the multipole expansion, and
smooth out the acceleration and the potential by means of the Hernquist \& 
Katz (1989) formalism. We adopt the opening angle $\theta = 0.8$,
and tie the gravitational softening parameter $\epsilon$ 
 to the particle number by looking at the mean inter-particle
separation at half-mass radius. In the simulations below  $\epsilon
 \approx 0.15 $.

\subsection{Description of the tests}

The temporal evolution of the system is shown in the various panels 
of Figs.~\ref{evol_vr_ad} and \ref{evol_ene_ad} 
which display the radial velocity and the specific internal 
energy, respectively. Each panel shows the variation of the physical
quantity under consideration (in suitable units) as a function of the
normalized radial coordinate at different time steps. 
These (in units of the dynamical time scale of the
system) are exactly the same as in Hernquist \&  Katz (1989).
  
The initial low internal energy is not sufficient to support the gas 
cloud which starts to collapse. Approximately after 
one dynamical time scale a bounce 
occurs. The system can be described as an isothermal core plus an 
adiabatically expanding envelope 
pushed by the shock wave generated at the stage of 
maximum compression. 
After about three dynamical times the system reaches virial equilibrium
with total energy equal to a half of the gravitational potential energy.

The temporal evolution of the potential and kinetic energies,
 and the total angular momentum is shown in Fig.~\ref{ene_mom_con}.
As pointed out by Nelson \& Papaloizou (1994) the energy conservation,
expressed as $\frac{\Delta E}{E}$ is $\approx 8\%$.
This uncertainty is caused by neglecting $\nabla h$ terms and using
 multiple time steps. 
In brief, when the number of active particle gets too small, this
 causes fluctuations that may affect the energy conservation.
Evolving the system with an unique
time step for all the particle lowers the uncertainty on the  energy
conservation down to about $5-6\%$. As far as the angular momentum is 
concerned, this is conserved within one percent.
Going back to the panels of Figs.~6 and 7 the present results agree
fairly well with the mean values of the Hernquist \& Katz (1989) models
provided that the effects of the different number of particles are
taken into account. 
For example the shock is located at the radial distance  $0.18 \leq r/R
\leq 0.25 $ in the models with age $t \approx 0.88$ (cf. the velocity panels
of Fig.~\ref{evol_vr_ad}).
The thermal energy slowly increases for $r/R \leq 0.1$ (cf. the panels of 
Fig.~\ref{evol_ene_ad}). Finally, at the time
of maximum compression ($t \approx 1.1$), the dynamical range in 
the time steps is $20:1$.
The scatter shown by the velocity and internal energy 
 in the post shock phases is caused by the relative small number
of particles, and  the use of the multiple time step scheme.
Similar scatter is present also in Steinmetz \& Muller (1993) and
Nelson \& Papaloizou (1994).

\begin{table}
\tabcolsep 0.10truecm 
\caption{CPU time percentage in the adiabatic collapse.}
\begin{tabular}{lcc} \hline
\multicolumn{1}{c}{Section of the code} &
\multicolumn{1}{c}{Hernquist \& Katz} &
\multicolumn{1}{c}{This work} \\
\hline
Gravitational computation& 43.28&42.63\\
SPH computation&27.63&26.37\\
Search of the nearest neighbors &24.43&26.22\\
Tree construction&4.50&4.28\\
Miscellaneous&0.16&0.50\\
\hline
\end{tabular}
\end{table}

\noindent
The above numerical tests have been 
performed on a Digital ALPHA OSF/1 workstation. It took about 5400 
seconds of CPU time for 1019 time steps.
It is worth comparing the CPU times required by
different sections of our code with 
those by Hernquist \& Katz (1989) . The comparison is summarized in Table~1. 
It appears that the fraction of CPU time spent in key sections of our
code agrees with those of the Hernquist \& Katz (1989) code.

\section{Dark matter}
We consider now a collapsing cloud of gas and dark matter
with  initial solid body rotation. 
The inclusion of dark matter is an important step, because it 
is known to play a key role in any modern theory of galaxy
formation and evolution (cf. Frenk et al. 1996, Persic \& Salucci 1997).
 
Initially, the dark matter particles have the same
spatial distribution of the gas particles. Both in fact are obtained
stretching an  initial regular cubic grid (Evrard 1988). Therefore,
dark matter and gas initially obey  the same density profile.

\begin{figure*}
\centerline{\psfig{file=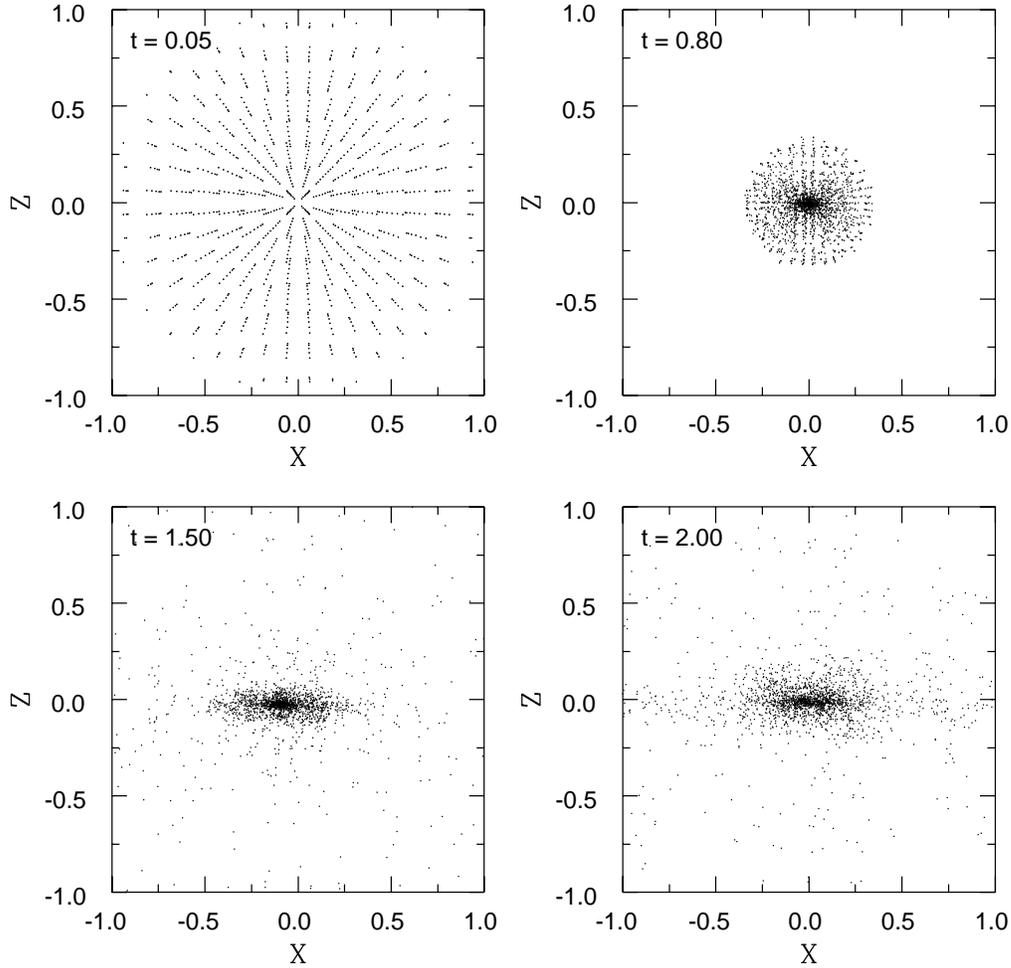,height=14cm,width=14cm}}
\caption{Adiabatic collapse: the evolution of the dark matter  
component as a function of time
in the X-Z plane. 
Time is in code units}
\label{ev_dark}
\end{figure*}

\begin{figure*}
\centerline{\psfig{file=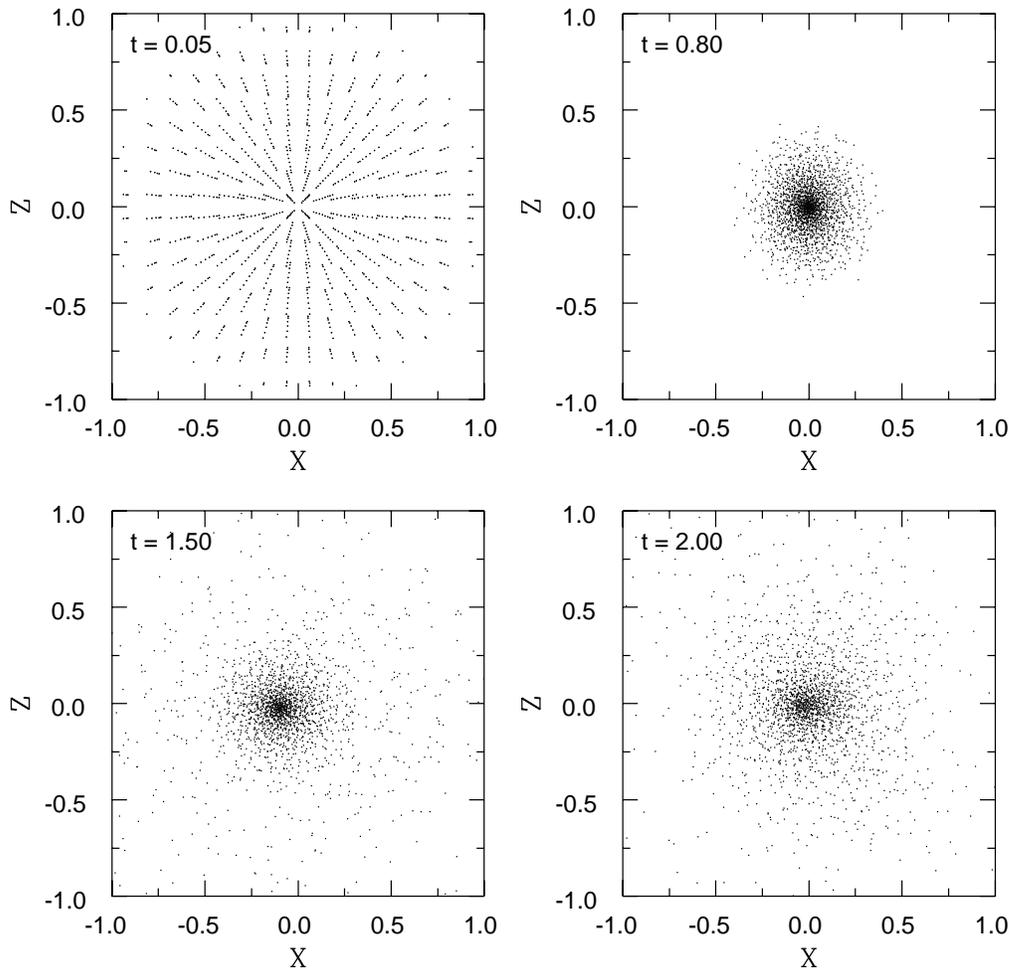,height=14cm,width=14cm}}
\caption{Adiabatic collapse: the evolution of the baryonic 
component as a function time in  
the X-Z plane. 
Time is in code units}
\label{ev_bary}
\end{figure*}

\begin{figure*} 
\centerline{\psfig{file=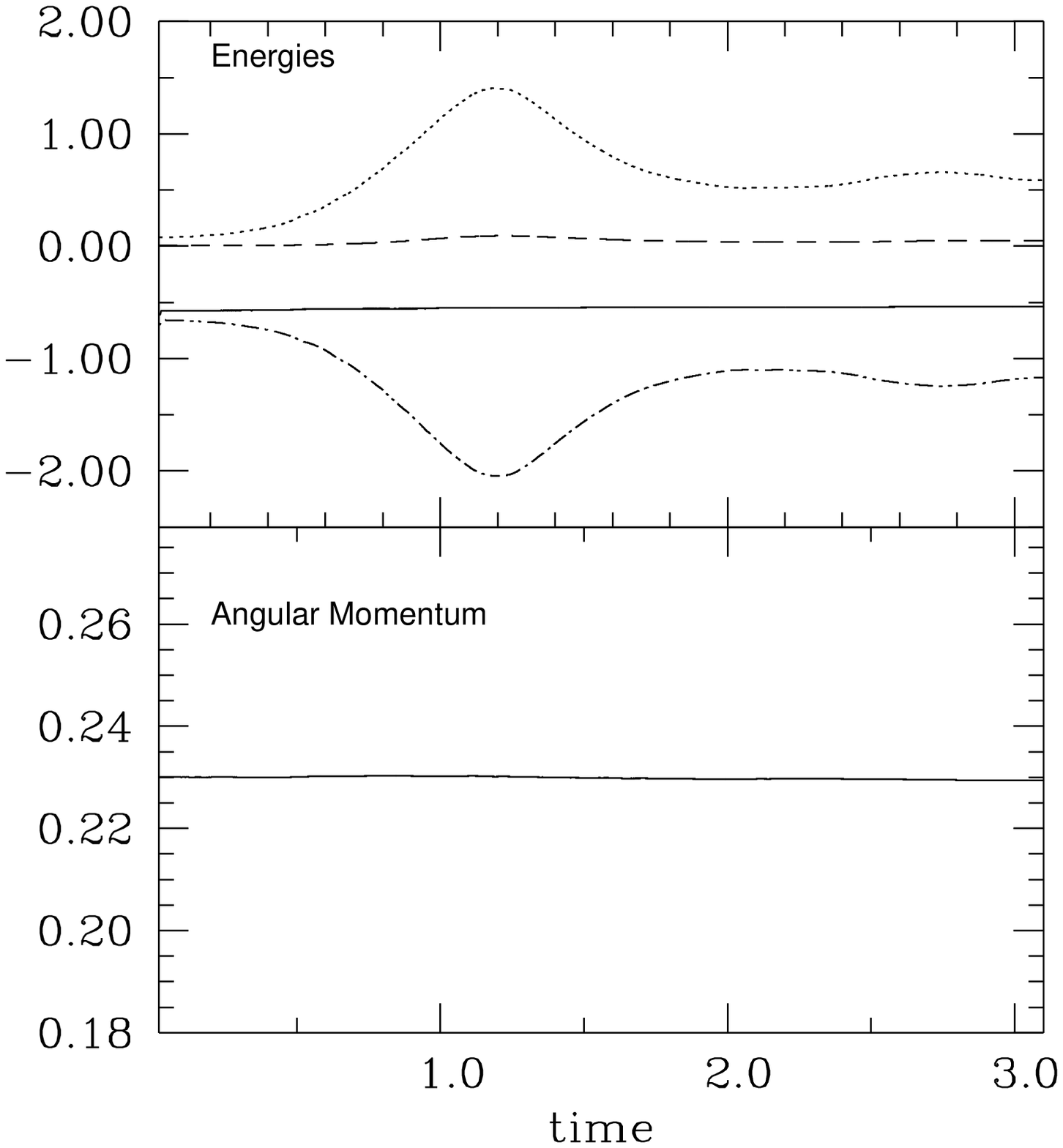,height=12cm,width=15cm}}
\caption{Collapse of a clump of gas and dark matter.  Upper panel: conservation of the potential (dashed-dotted line), 
kinetic (dotted line),  thermal (dashed line), and 
total energy (continuous line). Lower panel: conservation of the
total angular  momentum}
\label{ene_mom_con_dm} 
\end{figure*}

\noindent
This is a source of a technical difficulty with the tree-code algorithm
because when two 
particles occupy the same position, the tree-code is not able to distinguish
among them, when trying to divide the volume into elementary cells
each of which containing either only one particle or no particles at all.
This difficulty does not of course occur with the binary tree-code 
algorithm (cf. Benz et al. 1989).
To handle this problem within a single tree-code scheme, we force the tree-code
to consider a subdivision as an elementary  cell also when two particles 
are enclosed
provided that they are of different nature. This implies
that every particle needs an additional flag specifying  the species.

The system under consideration simulates a spherical galaxy whit total mass 
and radius of $10^{12} M_{\odot}$
and  100 kpc, respectively. In addition, the system is supposed to be made
of equal numbers (2176) of particles in form of baryons   and dark matter
but with different individual mass such that the total mass fraction of
baryons is 0.1 the total mass of the galaxy.

We adopt an initial solid body rotation around 
the Z--axis with angular velocity $\omega \approx 0.5$ which corresponds 
to a frequency of about 1 complete rotation in 10 free--fall 
time scales. So the effects of rotation 
get sizable during the collapse. This rotation corresponds to adopting the
dimensionless spin parameter 

\[
\lambda = \frac{J |E|^{1/2}}{G M^{5/2}} = 0.08
\]

is  somewhat larger than
the value expected from the tidal torque in the hierarchically
clustering universe theories, i.e. $\lambda \approx 0.05$ (White 1984,
Steinmetz \& Bartelmann 1996).

The gravitational softening parameters $\epsilon$ for gas and 
dark matter are in code units (see below)
 0.002 and 0.010  to which 200 pc and 1 kpc
correspond, respectively. They have been fixed on the basis of the
Evrard (1988) relation

\begin{equation}
\frac{G m}{\epsilon} \ll  \frac{G M}{R} .
\end{equation}

\noindent
The parameter $\epsilon$ is kept constant in time for each individual 
particle. However if in a cell two particles of different nature (gas
and DM) co--exist, the softening parameter is the mass weighted mean
value. 

Tests similar to those described below can be found in Navarro \& White 
(1993) to which our results can be compared. In this context, the 
most salient result of the Navarro \& White (1993) calculations is
the different evolutionary behaviour of the dark matter
 and baryonic components.
The same results are recovered here as shown in Figs.~\ref{ev_dark} and 
\ref{ev_bary}, respectively.
In our models, however,  the difference  is less pronounced because
of the lower initial spin as compared to  $\lambda = 0.10$ in Navarro \& 
White (1993). In brief, dark matter 
gets a more   flattened spatial distribution than the baryonic component.
In fact equidensity contours of the baryons must coincide with
equipotentials which are much rounder than contours
of equal total density.
It is worth recalling that observational hints for flatter 
dark halos have been found in 
galaxies like NGC~5907 (Sackett et al. 1994),  NGC~4244 (Olling 1996),
and NGC~891 (Becquaert \& Combes 1997).

\section{Non adiabatic processes}

As far as the evolution of the gaseous component is concerned, 
all non adiabatic
processes but artificial viscosity  are of paramount importance.
They drive in fact the energy and momentum loss or gain of these "particles".
Since the final target is to follow the evolution
of a real galaxy, processes like cooling and heating, star formation,
and feed-back play a dominant role, and must be included from the very
beginning in order
to obtain results worth to be compared with observational data. 
Unfortunately, some of these
processes are not yet well understood. The physics
of star formation, in particular,  is far from being assessed
so that parameterizations of this basic process are imposed.

\begin{figure*}
\centerline{\psfig{file=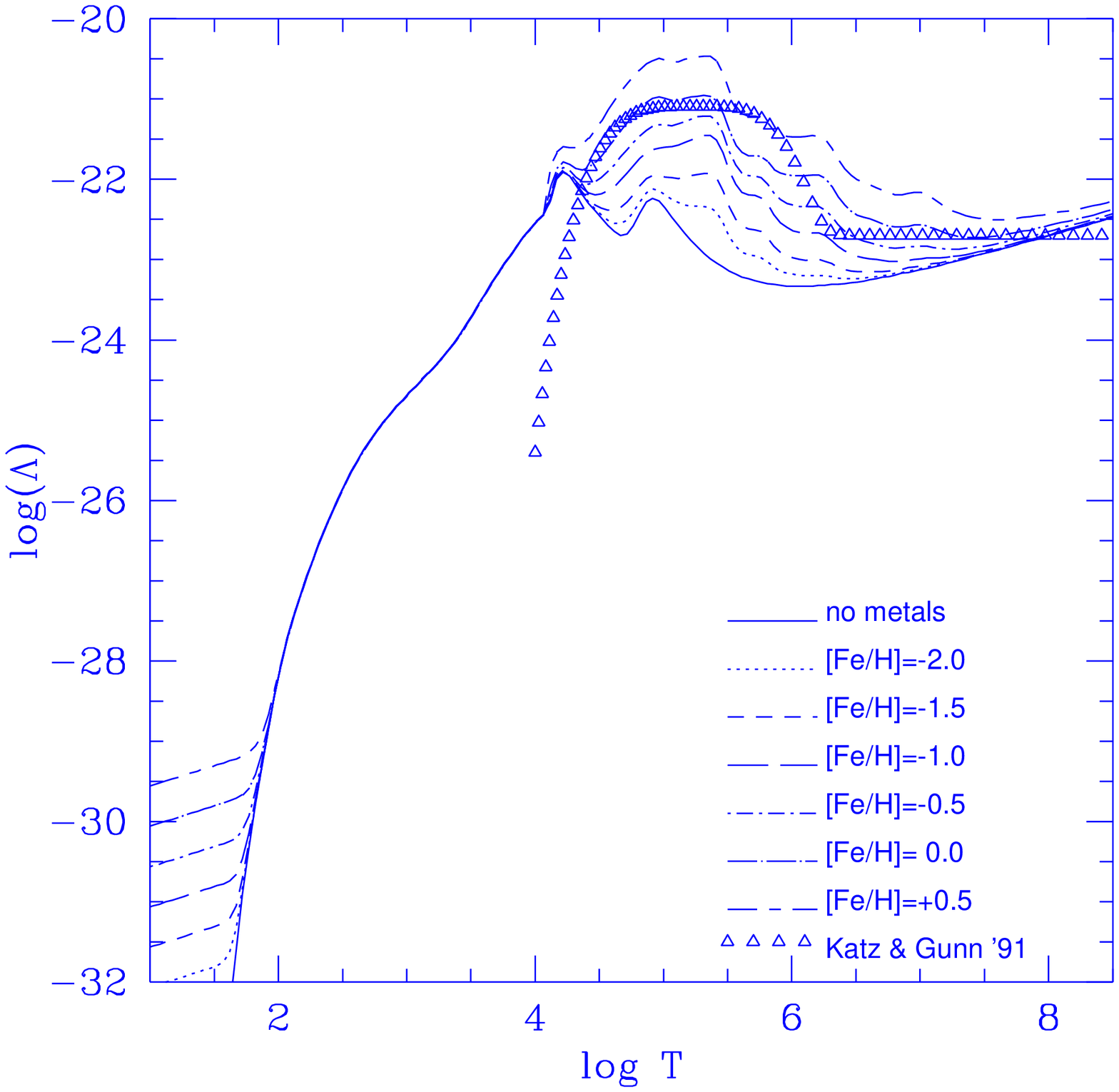,height=10cm,width=12cm}}
\caption{Net cooling rate $\Lambda_C/n^2$ in ergs 
$\rm cm^{-3}~ s^{-1}$ as a 
function of temperature. $n$ is the number density of particles. The
cooling rate is  from Sutherland \& Dopita (1993) for $T \geq 10^4$, 
Hollenbach \& McKee (1979) and Tegmark et al. (1996) for 
$100 \leq T \leq 10^4$, and Caimmi \& Secco (1986) and Theis et al. (1992) 
for $T \leq 100$ K. 
Each curve corresponds to a different
metallicity as indicated. 
For comparison the  cooling rate by Katz \& Gunn (1991)  is superposed 
(open triangles) }
\label{cool_rate}
\end{figure*}

In this section we present in some detail our assumptions
concerning  thermal conduction and cooling processes of the gaseous
component, which are known to regulate  the energy budget
of a fluid element. They are included in the energy equation as 
source terms.

\subsection{Thermal conduction}

Thermal conduction is calculated according to  the formalism 
developed by Monaghan \& Lattanzio (1991)

\begin{equation}
\frac{1}{\rho} \vec \nabla \cdot (\rho q \vec \nabla u)  .
\end{equation}

\noindent
 The  translation of it into the
 SPH language is

\begin{equation}
\sum_{j=1}^{N} m_j \frac{(q_i - q_j)(u_i - u_j) \vec r_{ij} \cdot
 \vec \nabla W(\vec r_{ij},h)}{\rho_{ij} (r_{ij}^{2} + \eta^{2})},
\end{equation}

\noindent
where $q$ is the thermal conductivity function. This is in turn evaluated as

\begin{equation}
q_{i} + q_{j} = g h (c_{ij} - 4\cdot  \mu_{ij}),
\end{equation}

\noindent
In the above relations,  $\eta$ is a parameter  securing that the
denominator  of relation (25) remains different from zero, and
and $g$ is another suitable parameter set equal to 0.25.

Thermal conduction is expected to be effective in situations
of strong shocks, and/or   "wall heating" when two flows 
or two fluid elements collide. Thermal conduction is a source of heating
in the energy equation (see below).

\subsection{Cooling}
Radiative cooling is the crucial mechanism responsible
for condensation and collapse of baryonic gas into galaxies and,
inside galaxies, for the occurrence of  star formation.
 Cooling processes are rather well known, and have already 
been included in current models of galaxy formation and 
evolution albeit in different detail. See  for instance 
Katz, Weinberg \& Hernquist (1995) for a very accurate treatment 
of cooling processes of different nature.
Among the various cooling agents we recall: (i) the radiative cooling 
by atomic and molecular processes; (ii) the inverse Compton mechanism 
in presence of microwave background radiation.
\littleskip

{\bf Radiative cooling}. 
The most commonly used radiative cooling functions are those by Katz
\& Gunn (1991) because they are analytical and
continuous with their first derivatives. These radiative cooling functions are
calculated assuming the typical  helium abundance  by mass Y=0.25. 
They do not contain, however, the dependence on metallicity
 which in contrast is expected to be important during the evolution of a 
galaxy, as a natural consequence of the  chemical enrichment. 
To cope with this drawback of the Katz \& Gunn (1991) cooling functions,
we adopt here those elaborated by Chiosi et al. (1997). In brief, 
these cooling function $\Lambda_{C}(u_i,\rho_i,Z_i)$ are derived 
from literature data,  and include different radiative
processes. For temperatures greater than $10^4$ K they lean on
the Sutherland \& Dopita (1993) tabulations for a plasma under equilibrium 
conditions and  metal abundances  [Fe/H]=-10 (no metals), -3, -2, -1.5, -1, 
-0.5, 0 (solar), and 0.5. 
 For temperatures in the 
range $100 \leq T \leq 10^4 $ the dominant source of cooling is the $H_2$ 
molecule getting 
rotationally or vibrationally excited through a collision with an
$H$ atom or another $H_2$ molecule and decaying through radiative 
emission. The data in use have been derived from the analytical expressions of 
Hollenbach \& McKee (1979) and Tegmark et al. (1996). 
Finally, for temperatures lower than 100 K, starting from the relation of
Theis et al. (1992) and Caimmi \& Secco 
(1986), they corporate the results of Hollenbach 
\& McKee (1979), and Hollenbach (1988) for CO  as the dominant coolant. 
The following analytical relation 
 in which the mean fractionary abundance of CO
is given as a function of [Fe/H], is found to fairly represent the 
normalized cooling rate (i.e. $\Lambda_C(CO)/n^2$ with $n$ the number density
of particles)

\begin{equation}
 { \Lambda_{C}(CO) }  =  1.6 \times 10^{-29} 
                      10^{([Fe/H] -1.699) }  T^{0.5}~~\rm erg~cm^3~s^{-1}  . 
\label{eq_14}
\end{equation}
The normalized  cooling rate (in units of $erg~cm^3~s^{-1}$) 
over the whole temperature range is shown 
in Fig.~\ref{cool_rate}, in which the effects of the metallicity are clearly 
visible.
It is worth mentioning that no re-scaling of the cooling rate from the various
sources has been applied to get the smooth curves shown in  
Fig.~\ref{cool_rate}. 
For the purposes of comparison, we display in Fig.~11 the cooling rate
of Katz \& Gunn (1991). The major point of disagreement is soon evident
because their cooling rate which by definition is for a zero-metal
composition actually lies well above the zero-metal curve of
Sutherland \& Dopita (1993). This implies that Katz' \& Gunn's (1991) 
cooling is unphysically more 
efficient in situations of nearly zero metallicity, 
such as the protogalactic phase.

\begin{figure*}
\centerline{\psfig{file=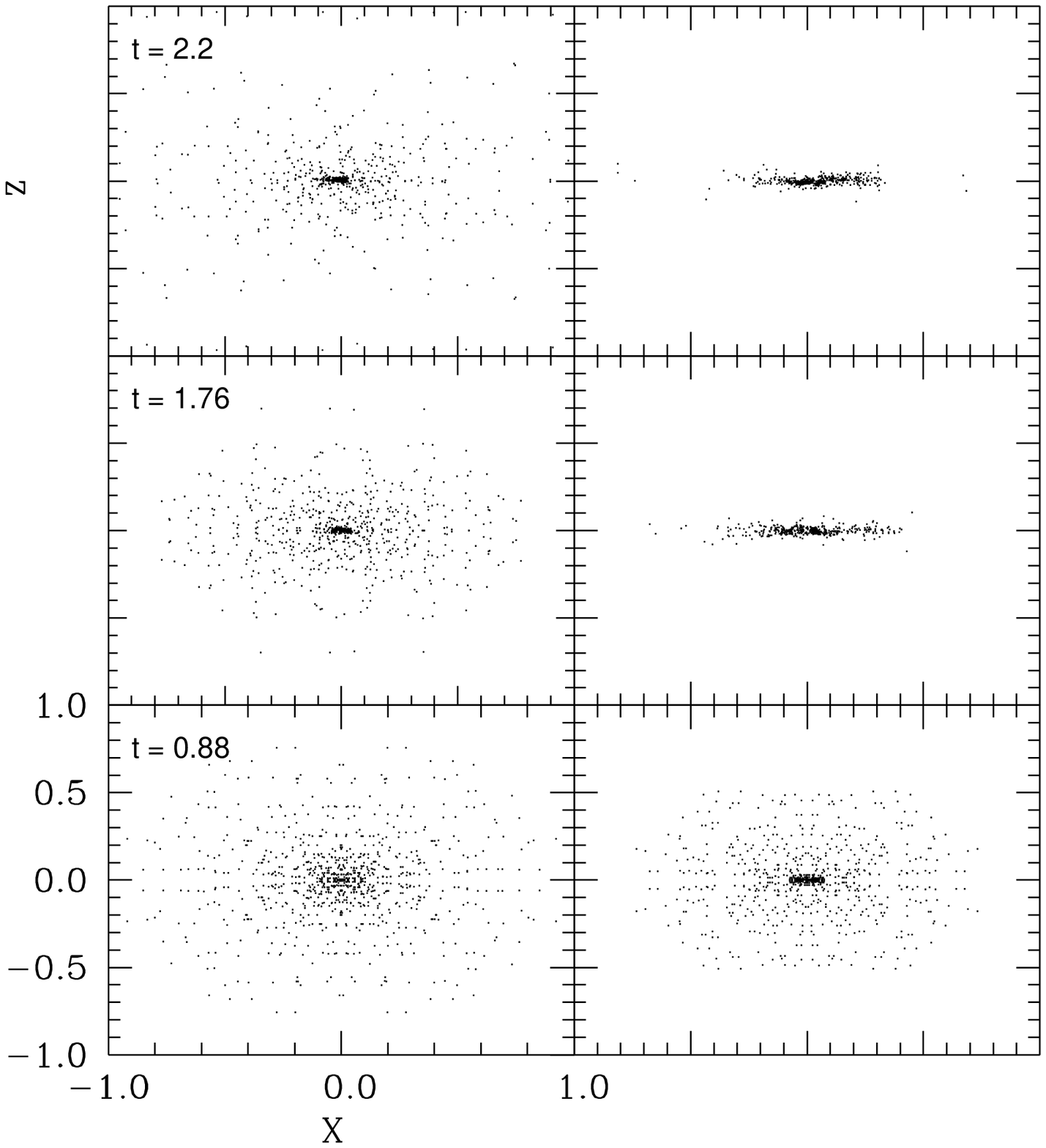,height=14cm,width=14cm}}
\caption{Testing cooling: the  final distribution in $X-Z$ plane for
gas particles in  simulations with different cooling
prescriptions. The right
panel is for the cooling rate by Katz \& Gunn (1991), whereas
 the left panel is for the cooling prescription described in the text }
\label{mod_cool_xz}
\end{figure*}

\begin{figure*}
\centerline{\psfig{file=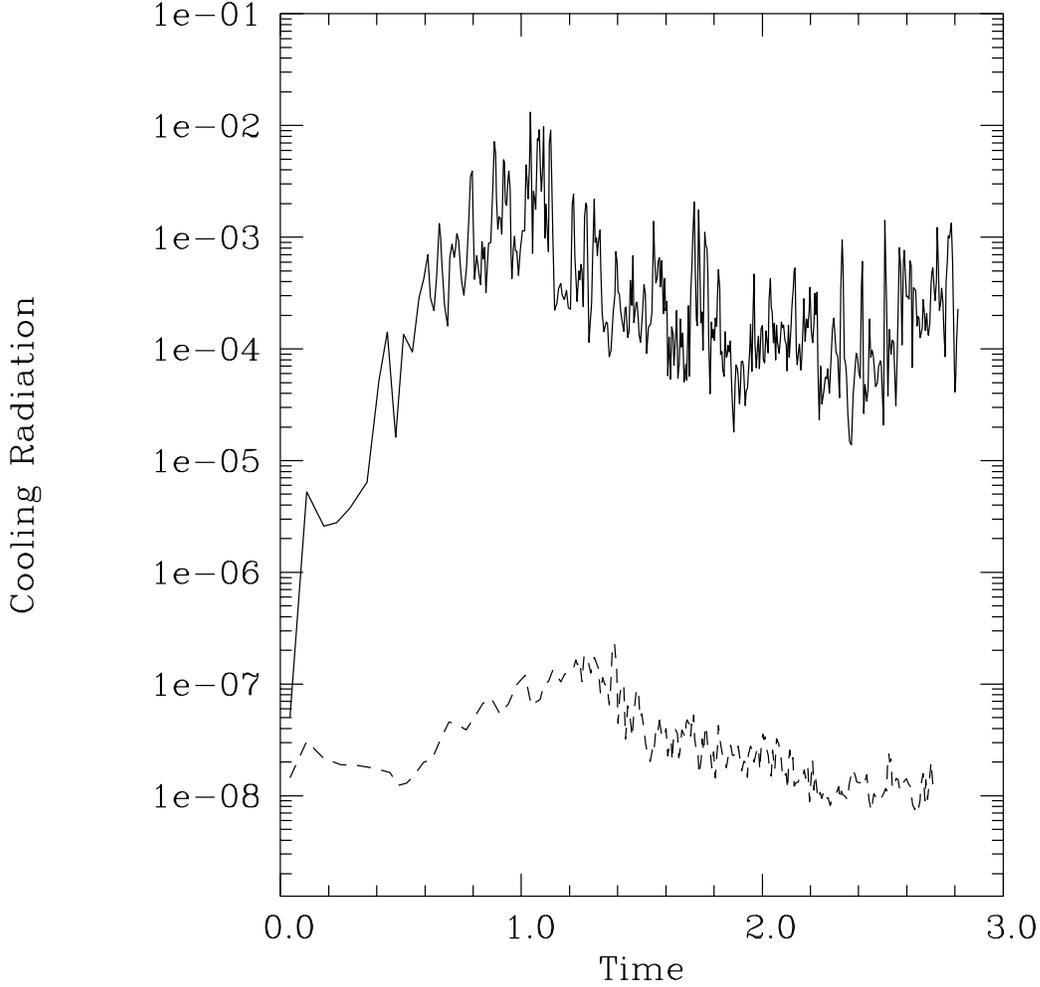,height=14cm,width=14cm}}
\caption{Cooling radiation (in code units): the solid line is with the 
cooling rate
by Katz \& Gunn (1991), whereas the dashed line is for the cooling rate used in this
paper }
\label{cool_rad}
\end{figure*}

\begin{figure*}
\centerline{\psfig{file=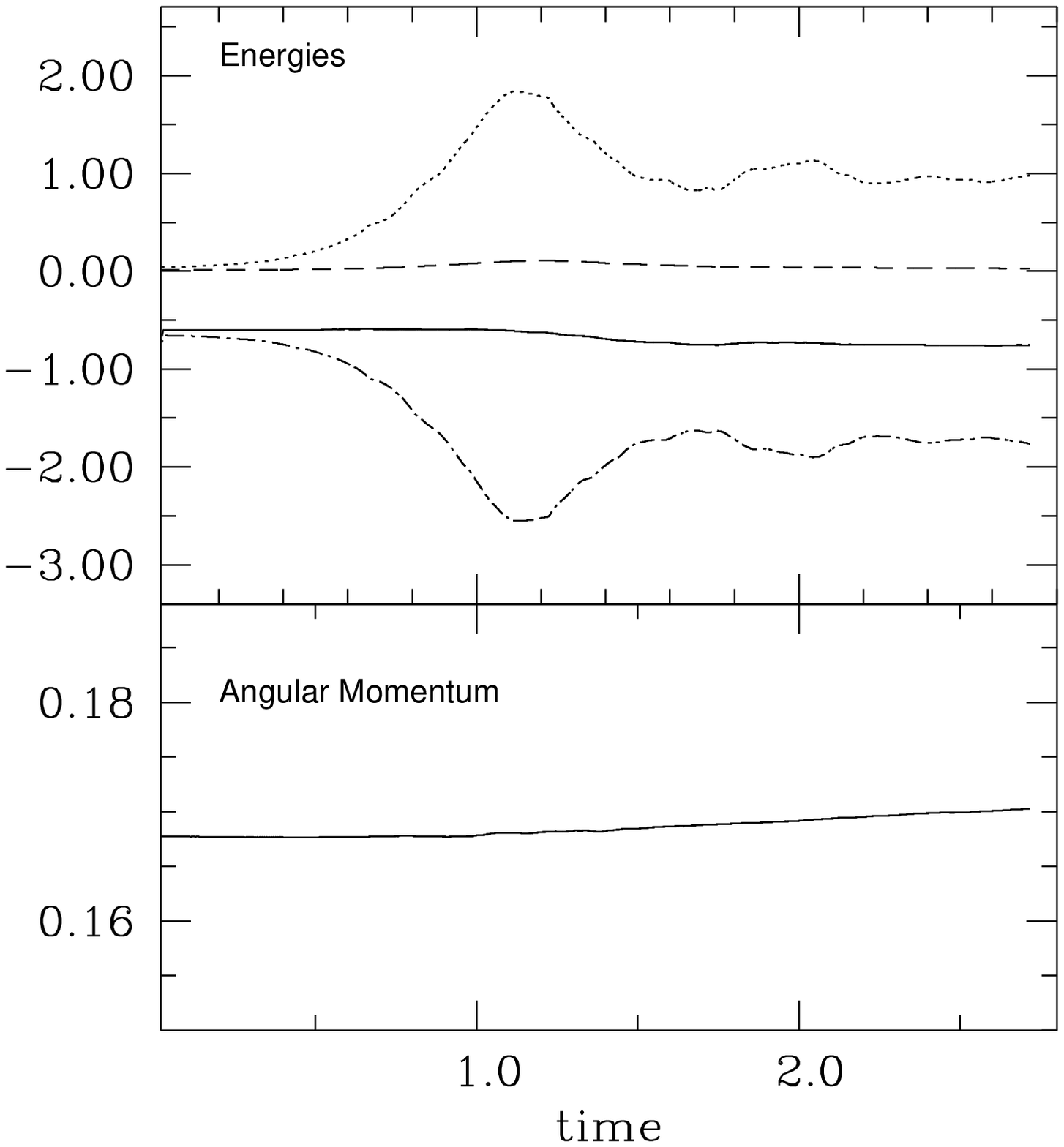,height=14cm,width=14cm}}
\caption{Conservation of energy and angular momentum in a collapsing galaxy
with the cooling rate adopted in this paper. The model is the same as in the left
panel of Fig.13  }
\label{cons_ene_mom_cool}
\end{figure*}

\littleskip

{\bf Inverse Compton}.
We do not take into account inverse Compton cooling because of
its strong dependence on redshift (Ikeuchi \& Ostriker 1986).
During the protogalactic phase of the evolution of a galaxy,
from which we start our simulations,
the inverse Compton cooling has already got much lower than the radiative
cooling, so that it can be neglected.

\littleskip

Although the  cooling rate has been derived over a wide range
of temperatures from a few degree  to above $10^8$, in reality
only the portion above a suitable temperature $T_{min}$ can be used in
model calculations because of the maximum resolution 
achievable in the  N-body simulations.
This  limit is derived from imposing
that the Jeans mass does not fall below a critical value, which is
conventionally assumed to be the mass of four gas particles (cf. Katz \& Gunn
1991). The limit temperature is given by the Bonnor-Ebert expression
(Ebert 1955, Bonnor 1956)

\begin{equation}
T_{min}=0.89553\left [\left (\frac{4m_p}{\rho_g}\right )^{2/3}-
\left (\frac{\pi \epsilon_g^3}{3} \right )^{2/3}\right ]  
\frac{\mu m_H G}{k}\rho_g,
\end{equation}

\noindent
where $m_{p}$ is the mass of a baryonic particle, $k$ is the Boltzmann 
constant, $\mu$ and $\epsilon$ are the mean molecular weight and the
gas gravitational softening parameter, respectively.

The limit temperature $T_{min}$ is locally computed for every gas particle. 
Finally, the cooling functions  are stored in the code as  look-up tables.

\subsection{Heating}
Heating of the gas is caused by many processes among
which we consider the thermalisation of the energy deposit by
supernova 
explosions (both Type I and II), stellar winds from massive stars, 
 the ultraviolet flux from
massive stars,  the cosmic ultraviolet flux,
and finally sources of 
mechanical nature.
\littleskip

{\bf Supernovae}. 
The rate of supernova explosions $R_{SNI,II}(t)$ over the time interval 
$\Delta t$ is calculated according to  standard prescriptions 
(cf. Bressan et al. 1994; Tantalo et al. 1996, 1997; Greggio \& Renzini 1983
for Type I SN in particular). 
Knowing the amount of energy released by each
SN explosion, the total  energy injection over the time
interval $\Delta t$ is the sum of  two terms, 
$E_{SNI}$ and  $E_{SNII}$,  of type

\begin{equation}
E_{SNI,II} = \int_{\Delta t} \epsilon_{SNI,II} R_{SNI,II}(t') dt'   ,
\label{eq_10}
\end{equation}

\noindent
where $R_{SNI,II}(t)$ is the number of supernovae  per unit time, and
 $\epsilon_{SNI,II}$ is the energy injected per SN explosion.
This formulation slightly differ from the one commonly adopted in
chemical models of galaxies, in which 
$\epsilon_{SNI,II}$(t) is the thermalisation law of SN remnants include of
the cooling processes (cf. Bressan et al. 1994, Tantalo et al. 1996, 
and Gibson 1995). In our scheme, first cooling effects are left to
the energy budget equation, second owing the limitations implicit in 
the N-body technique we are not yet able to resolve the star forming fluid 
to the level of individual stars but only to that of massive cluster-like
structures in which star can be formed according to a given initial mass 
function (see below). Therefore, following in detail the thermalisation law
of SN is impossible. A reasonable compromise is obtained by using sufficiently
small time steps (cf. also Chiosi et al. 1997 for a similar approach). 
The explicit formulations for $R_{SNI,II}(t)$ will be presented in the next 
section. 
\littleskip

{\bf Stellar winds}. The rate of energy injection by stellar winds is 

\begin{equation}
E_{W} = \int_{\Delta t} \epsilon_{W} R_{W}(t') dt'  ,
\label{eq_11}
\end{equation}

\noindent
where $R_{W}$ is the number of stars per unit time 
expelling their envelopes 
during the time interval $\Delta t$ and, in analogy with the SN remnants,
$\epsilon_{W}$ is the kinetic energy of  
stellar winds.
A losing mass star is expected to deposit  into the interstellar medium the 
energy

\begin{equation}
   \epsilon _{W0} = \eta \times 
      { M_{ej}(M) \over 2 } ({ Z \over Z_{\odot} })^{0.75} \times v(M) ^2  , 
\label{eq_12}
\end{equation}

\noindent
where $M_{ej}(M)$ is the amount of mass ejected by each star of mass $M$, 
$v(M)$
is the velocity of the ejected material,  and $\eta$  is an
efficiency factor of the order of 0.3 (Gibson 1994 and references therein).
The calculation of the rate $R_W(t)$ is postponed to the next section. 
\littleskip

{\bf Ultraviolet flux from massive stars.}
The rate of energy injection from the ultraviolet flux emitted by massive stars
is

\begin{equation}
E_{UV} = \int_{\Delta t} \epsilon_{UV} R_{UV}(t') dt'   ,
\label{eq_UV}
\end{equation}

\noindent
where $R_{UV}$ is the number of massive stars per unit time 
 whose mass
is the range $10~\div~120~ M_{\odot}$, and $\epsilon_{UV}$ is the amount of
ultraviolet energy emitted by each star. 
To calculate $\epsilon_{UV}$ the following procedure is adopted. We suppose 
that  massive stars are located on the zero age
main sequence, i.e. obeying well known mass-luminosity-effective temperature
relationships, $L/L_{\odot}(M/M_{\odot}$) and $T_{eff}(M/M_{\odot}$). In
principle, there should be an additional dependence on the chemical 
composition which, however, can be 
neglected here. The relationships $L/L_{\odot}(M/M_{\odot}$) and
$T_{eff}(M/M_{\odot}$) are taken from the library of stellar 
models/isochrones of Bertelli et al. (1994 and references therein). 
For the sake of simplicity we can approximate the spectral energy distribution
of any such stars as pure black body emission of given $T_{eff}$, for which
we can immediately estimate the fraction $F_{UV}(M)$ of flux emitted 
short-ward of 4000 \AA\ (our range for UV light) by a star of mass M.
The UV flux emitted by each star is 

\begin{equation}
 \epsilon_{UV} = F_{UV}(M) \times L(M)   ,
\label{UV}
\end{equation}
 
\noindent
where $L$ is the total luminosity of the  star.
Once again the calculation of $R_{UV}(t)$ is postponed to the next section.
\littleskip

{\bf Cosmic UV radiation}. As far as $UV$ radiation field is concerned, its 
effects have been thoroughly 
 investigated by Navarro \& Steinmetz (1996) who reached the 
conclusion that the cosmic $UV$ radiation plays some role
only when the gas has already  accreted onto the protogalaxy 
from the surrounding medium. 
Heating by cosmic UV radiation is not included in the present models.
In this context, it is worth recalling that all models in which 
the cosmic UV radiation is the sole source of heating face the so-called
{\em overcooling} problem (cf. Navarro \& Steinmetz 1996). 

\littleskip

{\bf Total radiative heating.} 
 The  total heating rate due to radiative
processes  $H_R(u_i,\rho_i,Z_i)$ to be included in the
energy budget equation (see below) is 

\begin{equation}
  H_R= { E_{SNI} + E_{SNII} + E_{W} + E_{UV}   
                     \over \Delta t }
\label{eq_13}
\end{equation}

{\bf Mechanical heating.}
In principle there are various  mechanical processes that
contribute to heat gas particles. Two of them have been
explicitly taken into account in the basic energy equation 
(\ref{eq_energy}). They have been shortly referred to as
$\Gamma_{i,M}$ and are general functions of type
$\Gamma_{i,M}(u_i,\rho_i,Z_i)$.

\subsection{More details on the energy equation}
At this stage of the analysis we can write down the detailed
expression of the energy equation in all its components. To this aim 
we need only to convert the radiative heating into the 
conventional SPH formalism

\begin{equation}
       \Gamma_{R,i}(u_i, \rho_i, Z_i) = \sum_j H_{R,j} \times 
            {m_j \over \rho_j} W(i,j) 
\end{equation}

\noindent
with obvious meaning of the symbols.
The final energy equations is
\begin{equation}
 {du_i \over dt} = \Gamma_{M,i} +
           \Gamma_{R,i}(u_i,\rho_i,Z_i)   -
           {\Lambda_{C,i}(u_i,\rho_i,Z_i) \over \rho_i }
\end{equation}
\noindent
with
\begin{equation}
         \Gamma_{R,i}(u_i,\rho_i,Z_i) = H_{C,i}(u_i,\rho_i,Z_i) +  
              H_{R,i}(u_i,\rho_i,Z_i) 
\end{equation}

\noindent
where $\Gamma_{R,i}$ the sum of thermal conductivity and all
radiative heating sources. The energy $u_i$ is per unit mass,
whereas all other quantities are per unit mass and time.

The functional dependence of the cooling term imposes an implicit
solution of the energy equation. This is performed 
using an hybrid scheme, i.e. a combination
of the Newton-Raphson and bisection methods. 
The associated second order updating of the thermal energy 
follows the technique of Hernquist \& Katz (1989).

Finally, it is worth mentioning that under high cooling efficiency 
 situations can be met in which the energy may become
negative. To cope with this unphysical result of mere numerical nature
we impose that a gas particle cannot lose more than a half
of its thermal energy per time step (Hernquist \& Katz 1989).

\section{Testing cooling prescriptions}
In this section we show an ideal galactic model in which no star formation,
no chemical enrichment, and no energy deposit from the various sources
are let occur. This model is meant to show the net results of different
prescriptions for the cooling rate.  The model has zero-metal content
and apart from the cooling rate, it is identical to that by 
Katz \& Gunn (1991) so that the comparison  
is possible. 

We simulate the collapse of a spherical galaxy with mass of
$10^{12} M_{\odot}$ and radius of 100 kpc. Like in the previous test 
the baryon fraction is 0.1, while the initial gas temperature is
$10^{4}~^{o}\rm K$. In this simulation, time, spatial coordinates, energy, 
and density are expressed in the following units: 
$\rm [time]=7.41\times 10^{8} year$, $\rm [space]=100~\rm kpc$,
$\rm [energy]=4.30 \times 10^{14} erg~gr^{-1}$, and
$\rm [density]=1.61 \times 10^{-25} gr~cm^{-3}$.
The temporal evolution of the models has been
followed up to the age of 2.2 dynamical times ($\ approx 1.6 Gyr$). 

The results are schematically presented
in Figs.~\ref{mod_cool_xz} and \ref{cool_rad}. Fig.~\ref{mod_cool_xz}
 shows the distribution 
of particles on the $X-Z$ plane at three different ages (in code units) for 
our cooling rate, left panels, and Katz \& Gunn (1991), right panels,
whereas Fig.~\ref{cool_rad} compares the cooling radiation for the two 
assumptions for the cooling rate.

As the  zero-metal cooling rate of Sutherland \& Dopita 
(1993), on which our prescription stands for temperatures above $10^4$ K,
is much less efficient than that of Katz \& Gunn (1991), this easily explains 
why  
significantly longer time scales are need to form disk structures (cf.
Fig.~\ref{mod_cool_xz}), and the 
much lower amount of radiated energy (cf. Fig.~\ref{cool_rad}).

Finally, in Fig.~\ref{cons_ene_mom_cool} we show the conservation of total
energy and angular momentum as a function of time for the case with the new 
cooling rate. 
The small decrease of the total energy at increasing time
is caused by the lost of thermal energy radiated away by cooling
processes
(for comparison see the detailed trend of energies in Fig.~11).
Angular momentum is conserved within 5\% accuracy (see also Fig.~11 for
comparison).

\section{Going toward complete models: star formation, feed-back, and 
chemical enrichment}

Implementing star  formation into N-body simulations of
galaxies is a cumbersome affair, reflecting partly our poor
knowledge  of this fundamental process, and partly the limitations
still imposed by N-body technique itself.

To accomplish the task two basic steps are required: firstly we need to
determine the star formation rate (SFR), and secondly we must treat the
effects of the star formation on the interstellar medium
(the so-called feed-back).

\begin{figure*}
\centerline{\psfig{file=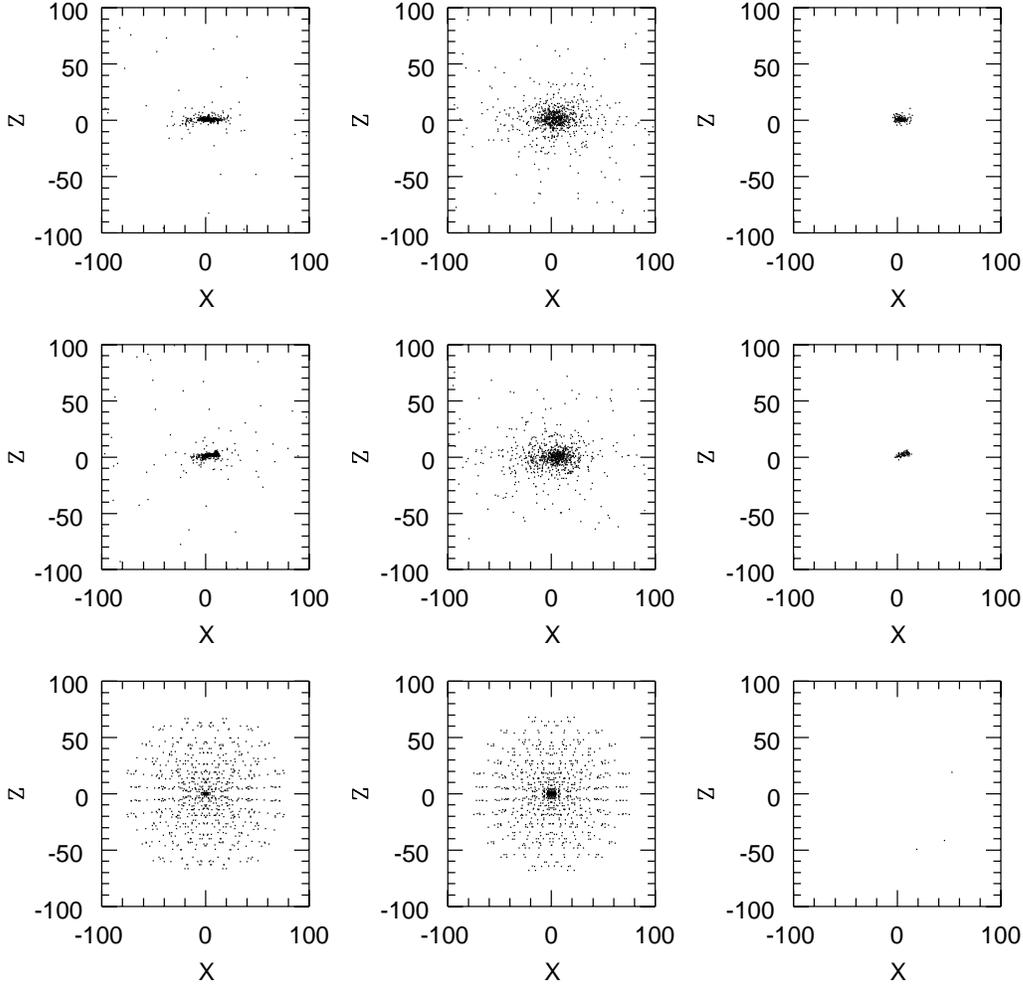,height=14cm,width=14cm}}
\caption{Simulation of a disk-like galaxy with no feed-back: 
the spatial distribution of 
 {\bf gas }(left panel) {\bf dark matter} (central panel) and 
{\bf stars} (right panel). Starting from the bottom the three panels of
each raw refer to ages of 0.5,  2.5, and 5 Gyr  }
\label{disk_xz}
\end{figure*}

\subsection{Star formation rate}
To include star formation in the N-body scheme two important informations are
needed: firstly we have to set suitable conditions under which stars can 
form,  secondly we must express the SFR law governing the efficiency
at which gas is turned into stars (cf. Steinmetz 1995 for a review on 
the subject).

{\bf When to form stars?}
There are at least three physical conditions to be met in order 
to activate the star forming process. First, 
a fluid element is eligible to make stars if it is part of  a convergent
flow, i.e. the particle velocity divergence must verify the 
 condition 

\begin{equation}
\vec \nabla \cdot \vec v_{i} < 0.
\end{equation} 

\noindent
Second, the fluid element must be Jeans unstable

\begin{equation}
\frac{h_{i}}{c_{i}} > \frac{1}{\sqrt{4 \pi G \rho_{i}}} .
\end{equation} 

\noindent
Finally, the  gas particle must remain cool

\begin{equation}
\tau_{cool} \leq \tau_{ff}.
\label{cond_3}
\end{equation}

\noindent
This last condition is usually translated into an overdensity criterion.
For instance, Navarro \& White (1993) uses the following expression

\begin{equation}
\rho_{i} > \rho_{crit},
\end{equation}

\noindent
where $\rho_{crit} = 7 \times 10^{-26} g~~cm^{-3}$.
This minimum threshold density actually holds only in the case of a single 
cooling function (no dependence on chemical composition). 
However, as star formation causes metal enrichment 
of the interstellar medium, the threshold moves towards lower
values at increasing metallicity. Because of this, we prefer to
 adopt  condition (\ref{cond_3}) which is of general validity.

\begin{figure*}
\centerline{\psfig{file=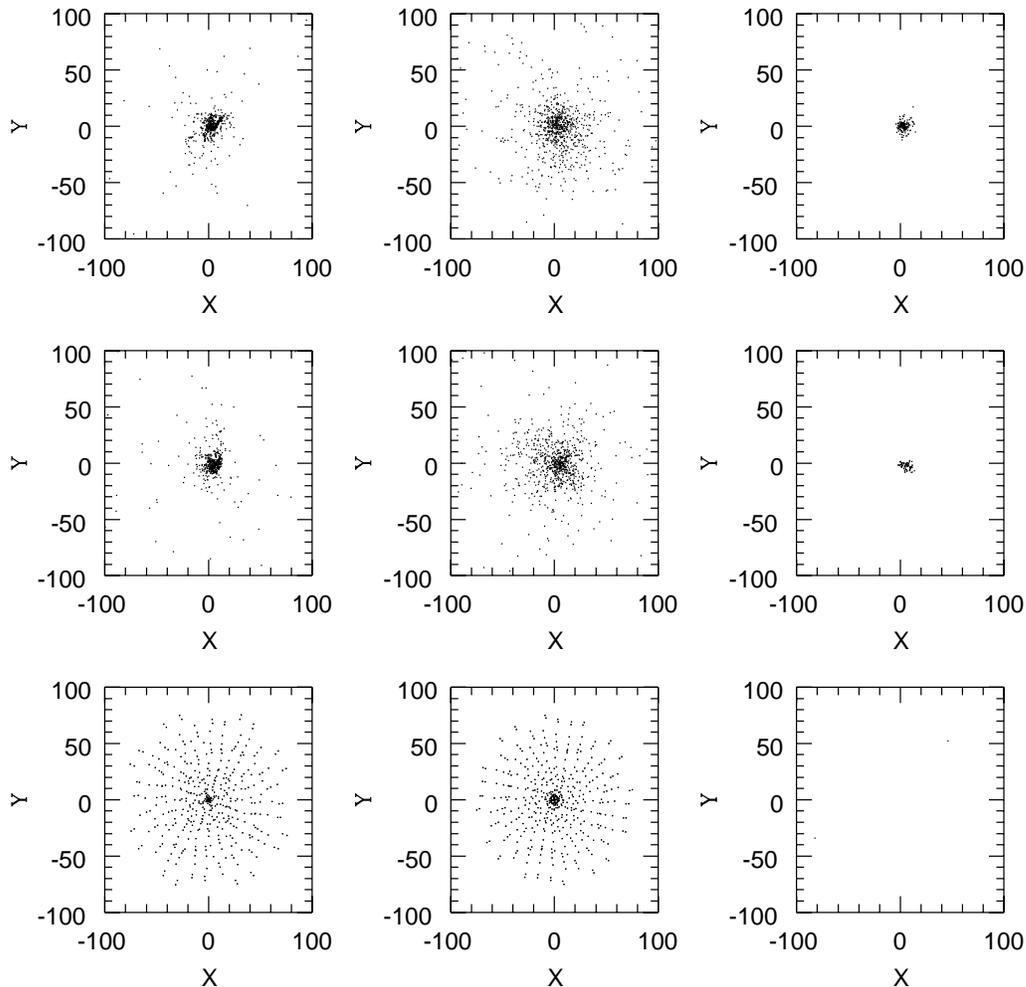,height=14cm,width=14cm}}
\caption{Simulation of a disk-like galaxy with no feed-back: 
the same as in Fig. 16 but onto the X-Y plane }
\label{disk_xy}
\end{figure*}

{\bf Star formation law.}
A very popular, empirically based, rate of star formation is 
the Schmidt  (1959) law, according to which

\begin{equation}
\Psi(t) = \frac{d \rho_{gas}}{dt} = - \frac{d \rho_{star}}{dt} = 
c_{\star} \frac{\rho_{gas}}{\tau_{ff}} ,
\label{sfr}
\end{equation} 

\noindent
where $c_{\star}$ is the specific efficiency (in most cases a free parameter).

In  dense regions (the ones prone to star formation), the cooling time
$\tau_{cool}$ is typically much shorter than the dynamical time
$\tau_{ff}$. Since $\tau_{ff} \propto \rho_{gas}^{-1/2}$, 
the SFR is proportional to $\rho_{gas}^{3/2}$, a result
that agrees with Schmidt's (1959) observational estimates.
In most numerical simulations, the specific efficiency is taken to be
$c_*=0.1$. 

\subsection{Star-like particles}
When star formation has started, at any time-step $\Delta t$ a star-like 
particle is created, whose mass is

\begin{equation}
m_{\star} = m_{gas} \times (1 - exp(-\frac{c_{\star} \Delta t}{\tau_{ff}})).
\end{equation}

\noindent
and the mass of the parent gas particle is consequently reduced.
The above relation simply follows from integrating equation (\ref{sfr}) 
over the time step $\Delta t$. 
The mass of the star-like particles obviously depends on the resolution 
(number of particles) of the simulation. 

The star-like particle is supposed not to immediately acquire 
its own individuality, but to leave the parent gas particle
only when the mass of 
this latter falls below a certain value (typically $50\%$
of the original mass). Furthermore, recurrent episodes of star 
formation within the same gas particle are possible, so that
gas is  depleted in discrete steps. Finally,  the
 star-like particle it treated as a 
collisionless object.
Therefore, the  element of fluid in which star formation occurs
is considered as a hybrid particle, whose collisional component experiences
both hydrodynamical  and gravitational forces, while its collisionless 
component feels only the gravitational field. 

To avoid non physical accelerations due
to the sudden decrease of the gas mass, particles in which star formation
is active are stored  in the lowest  time-bin.

When a star-like particle  leaves its parent, it inherits its position 
in the phase
space, and its gravitational softening parameter.
Resolution limitations  force on us to consider only four star formation
events per gas particle, like in Navarro \& White (1993).

\begin{figure}
\centerline{\psfig{file=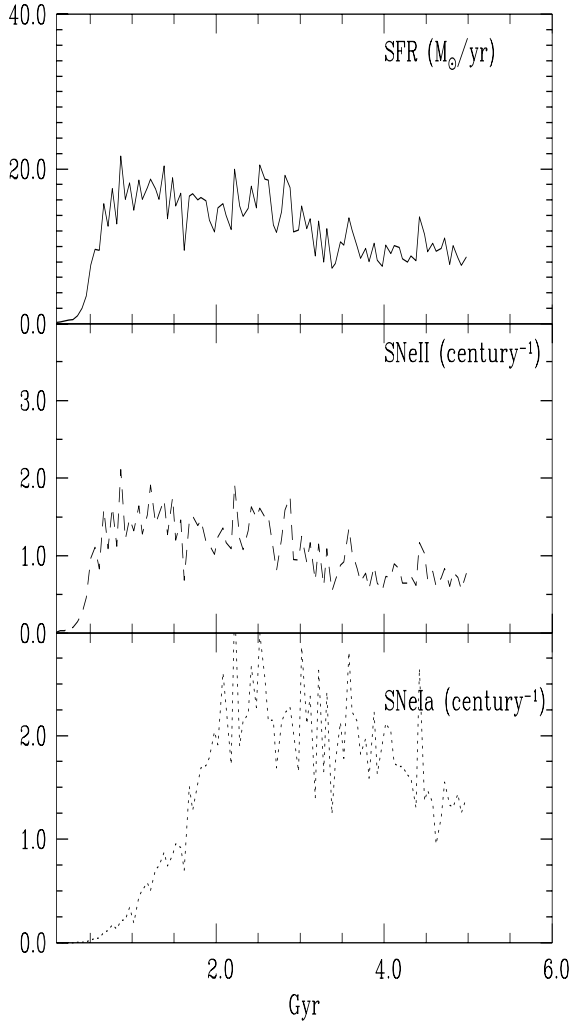,height=15cm,width=9.0cm}}
\caption{Simulation of a disk-like galaxy with no feed-back: the rates of
star formation (top panel) and supernova explosions (Type II middle panel,
Type I bottom panel) }
\label{sfr_no_feed}
\end{figure}

\begin{figure}
\centerline{\psfig{file=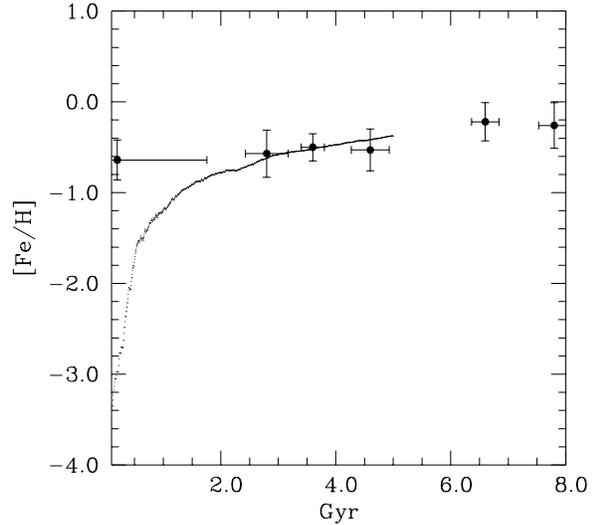,height=9cm,width=9cm}}
\caption{Simulation of a disk-like galaxy with no feed-back: 
the age-metallicity relation for the gas component in the galactic disk.
Filled circles show the age-metallicity relation of Disk stars in  the Solar
Vicinity by Edvardsson et al. (1993)}
\label{age_z_no_feed}
\end{figure}

\subsection{Feed-back}

To quantitatively evaluate the feed-back by all heating processes already
outlined in the previous section, we need to know the initial 
mass function, according to which the  newly borne stars
will distribute by number in each mass interval $dM$. 
For the sake of simplicity we adopt here
the classical  Salpeter (1955) law

\begin{equation}
\Phi(M)dM = A \cdot M^{-x}dM,
\label{imf}
\end{equation}

\noindent
where $x=2.35$ (the Salpeter value) and $A$ is the normalization constant.
This is 
derived from imposing that stars 
can be formed with masses in the range $0.1 \div 120 M_{\odot}$, and
assuming that the integral of eq. (\ref{imf}) over this mass range is 
equal to one. It is worth recalling that only stars more massive than 
about 0.9 $M_{\odot}$ will be able to pollute the interstellar medium 
over a time scale
shorter than the Hubble time (cf. Tinsley 1980, Matteucci 1997).
The normalization  constant is $A=0.06$. 
Other IMFs are already implemented into the code (Miller \& Scalo 1979).

{\bf Supernova rates }
In the standard scenario of stellar evolution (cf. Woosley 1986, Chiosi 
1986; Greggio and \& Renzini 1983) Type II supernovae occur in 
single stars more massive than say 8 $M_{\odot}$, whereas Type I (more
precisely Ia) supernovae occur
 in binary stars containing a re-juvenated white
dwarf brought to explosion via the C-deflagration mechanism by mass
accretion from the companion. Furthermore, Type II supernovae from
progenitors more massive than say 30 $M_{\odot}$ leave behind a black
hole (scarcely contributing to the enrichment of the interstellar medium in 
heavy elements), whereas Type II supernovae from progenitors in mass range
8 to 30 $M_{\odot}$ leave a neutron star and most effectively contribute
to chemical enrichment in heavy elements.
Type I supernovae have binary system progenitors with  typical total
mass in the  range 3 to 16 
$M_{\odot}$.  

\begin{figure*}
\centerline{\psfig{file=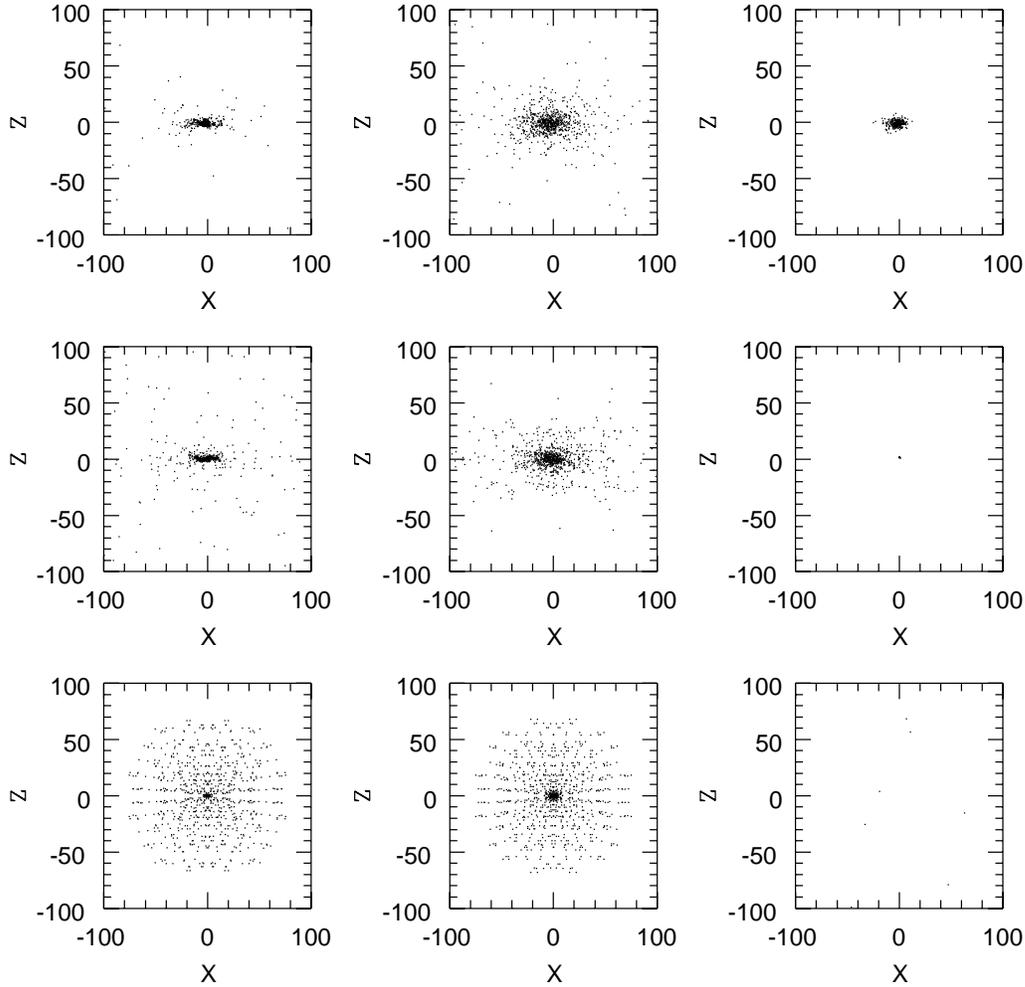,height=14cm,width=14cm}}
\caption{Simulation of a disk-like galaxy with full feed-back: 
the spatial distribution of 
 {\bf gas }(left panel) {\bf dark matter} (central panel) and 
{\bf stars} (right panel). Starting from the bottom,  the three panels of
each raw refer to ages of 0.5,  2.5, and 5 Gyr  }
\label{disk_xz_feed}
\end{figure*}

 To evaluate the rate of Type I  occurrence one has to know
the percentage of binary systems with respect to single stars. Let C be this 
fraction. Following Greggio \& Renzini (1983), 
the rate of Type I supernovae is

\[
R_{SNI} = C \int_{M_{B,inf}}^{M_{B,max}} \Phi_{B}(M_B) dM_{B} \times
\]
\begin{equation}
\int_{\mu_{inf}}^{1/2} 2^{\gamma + 1} (1 + \gamma) \mu^{\gamma}
\Psi[t - t_{(\mu M_{B})} ] d\mu .
\end{equation}

\noindent
In the above expression $M_{B}$ is the total mass of the binary system,
$\Phi_B(M_B)$ is the IMF of binary stars, which is 
assumed here to be the same as for single stars 
[eq.(\ref{imf})] for the sake of simplicity,
 $\mu = \frac{M_{2}}{M_{B}}$ is the fractionary mass
 of the secondary, which drives the evolution, and $\gamma$ is a 
coefficient, usually equal to 2. Finally, $M_{B,inf}$
and $M_{B,max}$ are the lower and upper limits for the total mass of the
binary stars, respectively, and $\mu_{inf}$ is the lower limit for the  
fractionary mass of the secondary. These masses are calculated with the
following receipt

\begin{equation}
M_{B,inf} = MAX(2 M_{2},2),
\end{equation} 

\begin{equation}
M_{B,max} = 8 + M_{2},
\end{equation} 

\begin{equation}
\mu_{inf} = MAX(M_{2}(t)/M_{B},1/2).
\end{equation} 

\noindent
Finally,  $\Psi(t)$ is the SFR  at the time equal to the difference 
between the current time and the lifetime of $M_{2}$. 
\littleskip

The total rate of Type II supernovae is

\begin{displaymath}
R_{SNII} =  \int_{M_{B,max}}^{120} \Phi(M) \\Psi(t_M) dM +
\end{displaymath}
\begin{equation} 
~~~~~~~~~~~~~~~~~(1-C)\int_{8}^{M_{B,max}} \Phi(M)\Psi(t_M) dM  
\end{equation}

\noindent
with obvious meaning of the symbols.
\littleskip
Finally, the constant $C$ above is calibrated looking at the rate
of Type Ia supernovae in spiral
galaxies (van den Bergh \& McClure 1994). $C$ turns out to be
$\approx 0.04048$. 
The contribution by supernovae explosions to feed-back has been also
included in the Tree-SPH model of the Milky Way by Raiteri et al. (1996).
\littleskip

{\bf Stellar winds and UV fluxes.}
The rates $R_W$ and $R_{UV}$, i.e. number of stars per unit mass and time
contributing to the energy injection by stellar winds and UV flux are
simply given by

\begin{equation}
R_W = R_{UV} = \int_{10}^{120} \Phi(M) \Psi(t_M) dM  .
\end{equation}

All mass-lifetime relationships entering the above calculations of the rates
$R_{SNi,II}$, $R_W$, and $R_{UV}$ are derived from the library of
stellar models/isochrones of Bertelli et al. (1994). Whenever
possible the effect of the chemical composition of the fluid element is taken
into account.
\littleskip

\begin{figure*}
\centerline{\psfig{file=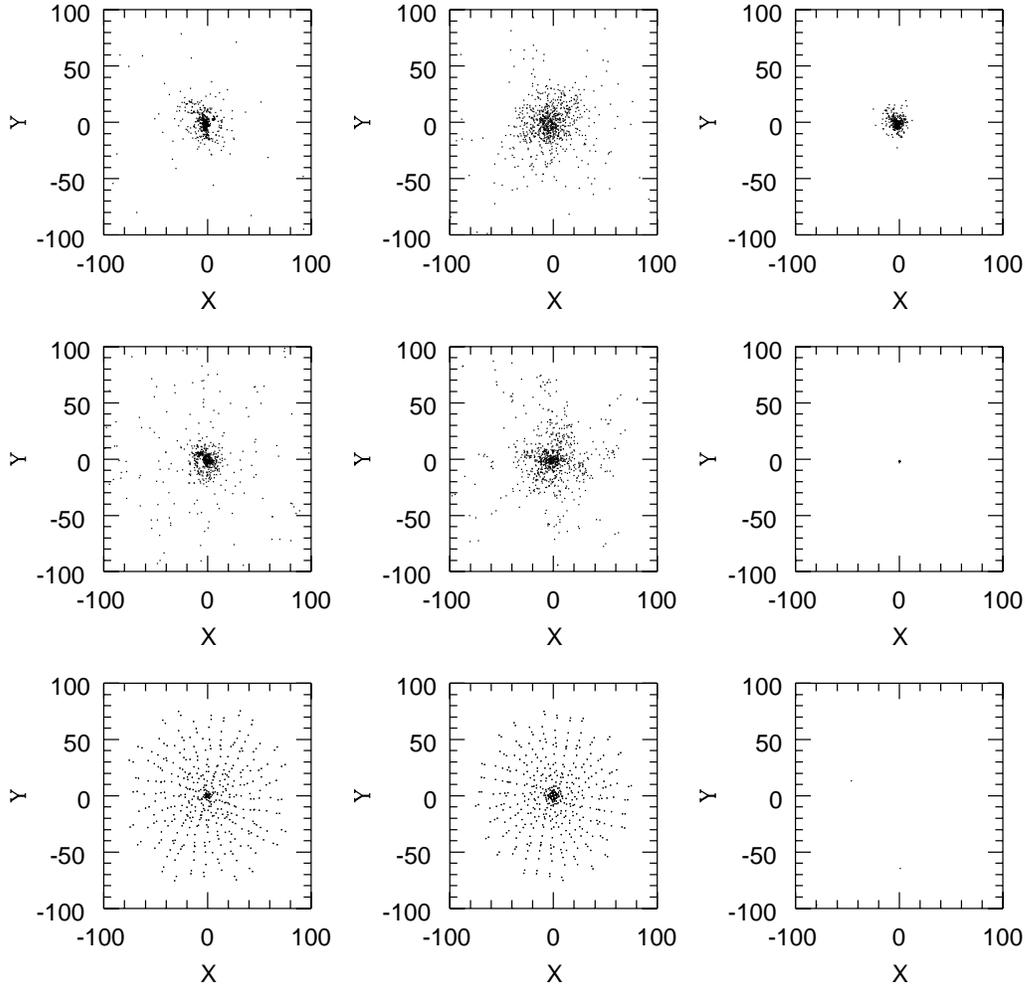,height=14cm,width=14cm}}
\caption{Simulation of a disk-like galaxy with full feed-back: 
the same as in Fig. 20 but onto the X-Y plane }
\label{disk_xy_feed}
\end{figure*}

{\bf Final remark.} Each supernova explosion  produces $10^{51}$ ergs of
energy which is
injected in the interstellar medium in form of kinetic energy.
A comparable amount of energy is supplied by a massive star during its
whole lifetime in form of stellar wind.
In the case of supernovae, 
only a small fraction of this energy is thermal (a few percent), 
and this small fraction is immediately 
(100 years is the typical time scale) radiated away
for instance in form of  $X$-rays emission. 
The effect of this energy input  on the velocity field of the
surrounding medium is expected to negligible. In brief, while 
the typical space
resolution of  current simulations is about 1 kpc or more, supernova shocks 
may affect  volumes of about 100 pc radius
 around the exploding star (Mazzali  \& Cappellaro, private communication)
Therefore it is not possible to follow the 
dynamics of such explosions down to the required resolution.
On the base of these considerations, the claim that some fraction of the 
supernova 
energy budget ($10^{-4}$ or so of the total) can modify the velocity
field of the neighboring volumes of our star-like particles
is not solidly grounded. 
We prefer  to simply assume that the cumulative effect of 
the supernova explosions increases the mean temperature and internal energy of the gas
on a scale larger than the size a fluid element.

\subsection{Chemical evolution}
Following , in our simulations,  
the chemical enrichment of the inter-stellar medium 
follows the prescriptions by Steinmetz \& M\"uller (1994).
It stands on the closed-box, instantaneous recycling
description (cf. Tinsley 1980), i.e.
the metal abundance $Z$ of a fluid element is computed as

\begin{equation}
\Delta Z_{i} = -Y_Z \frac{\Delta m_{i,gas}}{m_{i,gas}},
\end{equation}

\noindent
where $Y_Z$ is the so-called  yield per stellar generation. Given a certain
stellar nucleosynthesis scenario (cf. Matteucci 1997), $Y_Z$ can be
calculated a priori in a way mutually consistent with the assumed
IMF. Using the results of chemical models by Portinari et al. (1997)
a good assumption for yield per stellar generation is
$Y_Z = 0.005$.

Once metals are synthesized, they are spread according to the SPH
formalism:

\begin{equation}
<Z_{i}> = \sum_{j=1}^{N} Z_{j} \frac{m_{j}}{\rho_{j}} W(R_{ij},h).
\end{equation}

\noindent

Groom(1997) suggests that the chemical
enrichment of the interstellar medium is better described by
the diffusive equation

\begin{equation}
\frac{dZ}{dt} = - \kappa \nabla^{2} Z
\end{equation}

\noindent
where Z is the metallicity, and $\kappa$ the diffusion coefficient,
with $\kappa$ equal to 50 $\rm km~sec^{-1}~ 60~pc$.
The SPH translation, which requires second derivatives of the kernel,
is

\begin{displaymath} 
\frac{dZ_{i}}{dt} = - \sum^{N}_{j=1} \kappa m_{j} \left(
\frac{2}{\rho_{j}+
\rho_{i}}\right) \left( Z_{i} - Z_{j} \right) \times
\end{displaymath} 
\begin{equation}
~~~~~~~~~~~~~~~ | \nabla^{2}W_{ij}(|\vec{r}_{i}-\vec{r}_{j}|,h_{ij})|
\end{equation}

\noindent

Chemical enrichment is expected to occur on the same spatial scale
of the energy release of supernova explosions, i.e. about 100 pc,
which is much smaller than the spatial resolution of our model.

Although the use of $2^{nd}$ derivative of the smoothing function
may introduce spurious fluctuation in the resulting metallicity
implicit in equation (55), this
is less of a problem as far as the global enrichment in metals is concerned.

In fact, in semi--analytical models of chemical evolution of
disk--like
galaxies (Tinsley 1980, Carraro et al. 1997, 
Portinari et al 1997, and references therein)
the metallicity is know to quickly reach the metallicity yield 
of the contributing stellar population.

Therefore we expect the result not to critically depend on the
particular scheme used to evaluate the metallicity variation
with the SPH method.
This is implicitly confirmed by the agreement between theoretical
results
and observational data for the solar vicinity that are presented in
the section below.

We have adopted equation (55), and
converted the metallicity Z to [Fe/H]
by means of the relation

\begin{equation}
[Fe/H]= log(Z) + 1.739 
\end{equation}

\noindent
taken from Bertelli et al. (1994).

\section{The case of a disk-like galaxy}
The code described in the previous section has been used to study the
formation and evolution of disk-like galaxies. 
The initial model consists of a $10^{12} M_{\odot}$ spherical galaxy
enclosed within a radius $\rm R = 100$ kpc  and
containing  1000 particles of baryonic mass and 1000 particles
of  dark matter. We have assumed that the fractionary total mass of
 of baryons is 0.1 the total mass of the galaxy and 
 the spin parameter $\lambda=0.08$, which is in the range of current estimates
for disk galaxies.

\begin{figure}
\centerline{\psfig{file=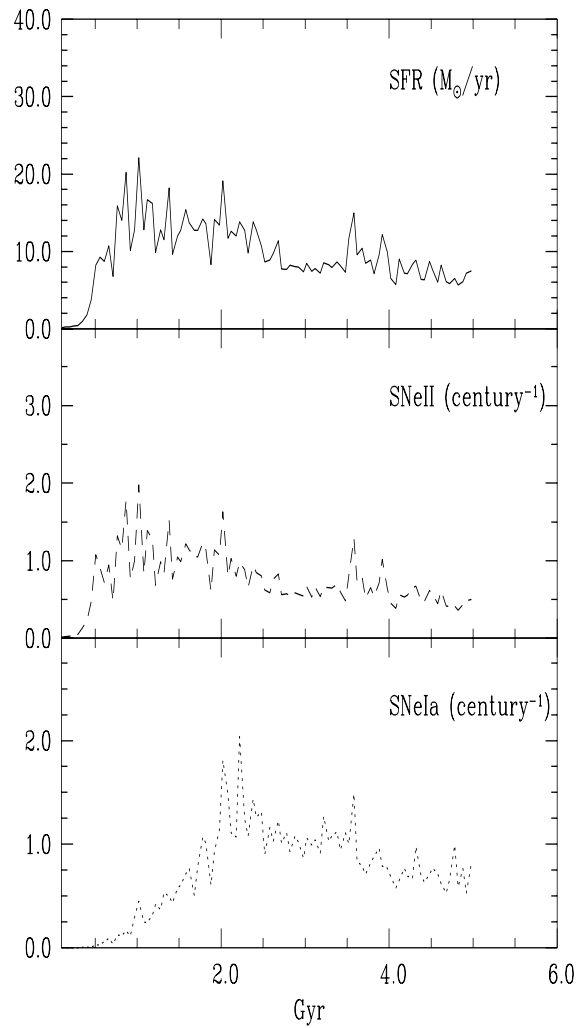,height=15cm,width=9.0cm}}
\caption{Simulation of a disk-like galaxy with full feed-back: the rates of
star formation (top panel) and supernova explosions (Type II middle panel, 
Type I bottom panel) }
\label{sfr_feed}
\end{figure}

\begin{figure}
\centerline{\psfig{file=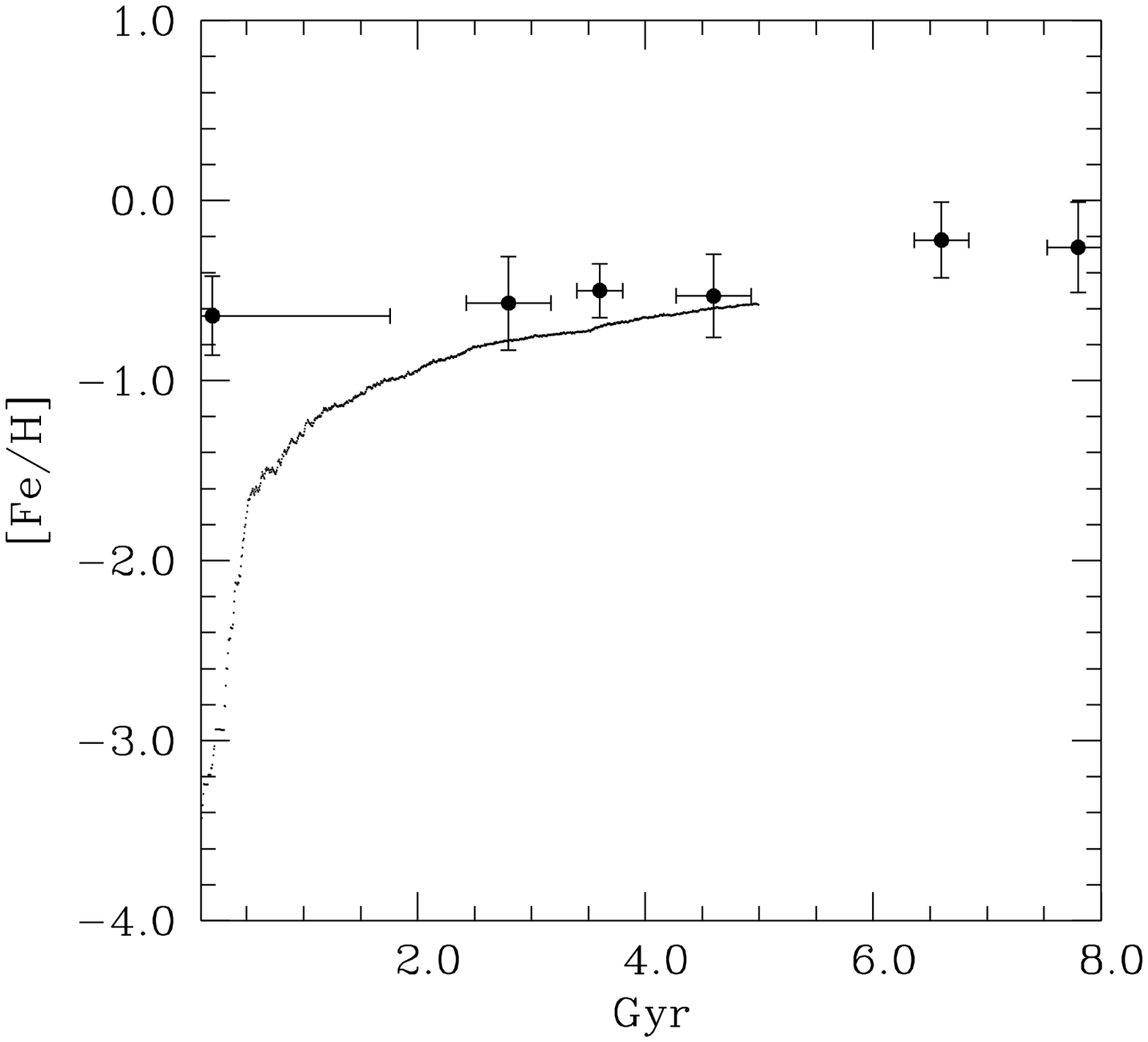,height=9cm,width=9cm}}
\caption{Simulation of a disk-like galaxy with full feed-back: 
the age-metallicity relation for the gas component on the galactic
disk. 
Filled circles show the age-metallicity relation for Disk stars in the
Solar Vicinity by Edvardsson et al. (1993)}
\label{age_z_feed}
\end{figure}

\begin{figure}
\centerline{\psfig{file=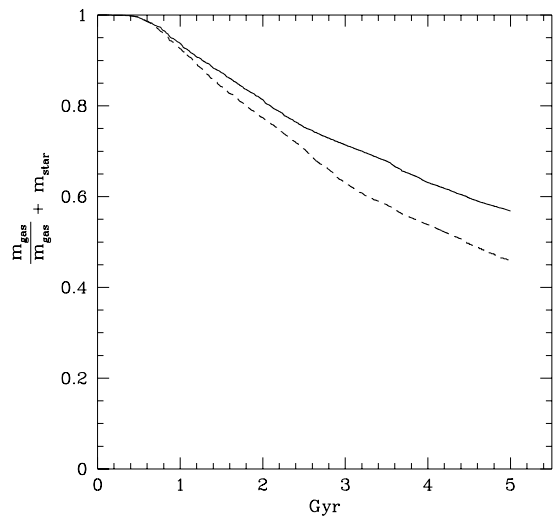,height=9cm,width=9cm}}
\caption{The gas fraction in models with no feed-back (dashed-line) and
models
with full fee-back (solid line)}
\label{gas_fraction}
\end{figure}

\subsection{No feed-back}

In Figs.~\ref{disk_xz} and ~\ref{disk_xy}
 we show the dynamical evolution of a disk-like galaxy in absence of feed-back.
Under the combined action of rotation and cooling, gas settles onto a disk
(cf. left panels of Figs.~\ref{disk_xz} and ~\ref{disk_xy}), and stars
 start to be formed. Remarkably, at the age of
about 5 Gyr a bulge-like structure made of stars is observed, whereas 
a disk-like
structure is not yet visible because of the insufficient resolution, and our
assumption that stars are considered as such when the gas fraction of each 
star forming unit has
fallen below 0.5 the total mass. We expect the disk to appear at later ages.
Like  the case of the collapsing galaxy without cooling and
star formation, a flattened  halo of dark matter can develop (cf. 
the central
panels of Figs ~\ref{disk_xz} and ~\ref{disk_xy} and Fig.\ref{ev_dark}).
Another important result of the new cooling rate is that gas particles survive 
in the halo for significantly long periods of time. 

The temporal evolution of the star formation and supernova rates 
is shown in Fig.~\ref{sfr_no_feed}. As expected the rate of star formation 
is low and nearly
constant with time. Furthermore, while Type II supernovae 
 closely follow the trend of the
star formation, Type I SN appear in significant proportions only  later than
about 1.5 Gyr due to the
rather long lifetime of their progenitors. 
Finally, in Fig.\ref{age_z_no_feed}, we show  the
age-metallicity relation for the gas component on the disk of the
galaxy and compare it with the age-metallicity relation for disk stars
in the solar vicinity by Edvardsson et al. (1993).  Despite the
crudeness of our modeling of chemical enrichment the agreement is
remarkable.
In particular at ages older than 2 Gyr, when the metallicity has
saturated to the yield.

\subsection{Complete feed-back}
The spatial evolution of the disk-like galaxy in presence of full feed-back
in shown in Figs.~\ref{disk_xz_feed} and \ref{disk_xy_feed}, which display
the gas (left panels), dark matter (central panels), and star particles
(right panels) at three different ages, namely 0.5 (bottom), 2.5 (middle)
and 5 Gyr (top). As expected the galaxy remains less spatially concentrated
owing to the energy input from the various sources and consequent heating.
This is particularly true for the gas and star particles, whereas no sizable
effect is seen on the dark matter component. The rates of star formation and
Type I and II supernova explosions as a function of time are shown in the 
three panels of Fig.~\ref{sfr_feed}. Although the global trend is as in the
 previous case, now the three rates are down by approximately a factor of two.
This can be easily understood as a result of heating which keeps the system
less spatially concentrated (lower density) thus immediately lowering the
rate of star formation and supernova explosions in turn.

In Fig.\ref{age_z_feed}, we show  the
age-metallicity relation for the gas component on the disk of the
galaxy and compare it with the age-metallicity relation for disk stars
in the solar vicinity by Edvardsson et al. (1993).  
The same remark made on Fig.~19  still applies.

Finally, in order to prove that
the mean density of the two model galaxies is actually different we plot in
Fig.~\ref{gas_fraction} the ratio $\rm m_{gas} /(m_{gas} + m_{stars})$
 as a function of the age. The dashed line is the galaxy with no feed-back
at all, whereas the solid line is for the galaxy with full feed-back.

\section{Conclusions and Future Perspectives}

In this paper, we have presented  a new Tree-SPH code suited to studies on the 
formation and evolution of galaxies.

Particular attention has been paid to include modern descriptions of non
adiabatic processes, such as cooling and heating by supernova explosions,
and stellar winds and UV radiation from massive stars,
 so that feed-back in the 
energy
balance is not a parameter but naturally follows from the assumed star
formation rate and initial mass function. Furthermore
we have followed the chemical enrichment of gas as the consequence of star
formation and include the dependence of the cooling rate on the gas chemical
composition (metallicity). These are points of major difference
and improvement with respect to similar codes in literature. 

Although the initial conditions are not tighten to any specific cosmological 
scenario, still the ones we have adopted lead to reasonable results. 
This point will be the subject of future improvements.

In addition to many classical tests aimed at checking the performance of the
code and its response to different physical assumptions, we have presented
the results for two disk-like galaxies with and without feed-back for
purposes of comparison both internal and with similar studies in literature.

Particularly relevant is the role played by the dependence of the cooling rate
on the chemical composition (metallicity), which owing to its strong impact
on the final results can no longer be neglected in simulations of galaxy
formation and evolution.

Work is in progress to extend these models to the case of elliptical galaxies
and to build the proper interface between this code and the 
spectro-photometric code of Bressan et al. (1994) so that the spatial history
of integrated spectra, magnitudes, colors, line strength indices etc. can be
derived together with the dynamical history for galaxies of different 
morphological type.

\section*{Acknowledgments}
G. C. deeply thanks W. Hillebrandt, S. White, E. M\"uller, and M.
Steinmetz 
of the Max Planck Institut f\"ur Astrophysik in
Garching for the warm hospitality and support during the six month
visit, that made possible to start this project. G.C. and C. L. thank
L. Danese, M. Persic, P. Salucci and R. Valdarnini for
encouragement and many useful conversations  during the year spent 
at ISAS, Trieste, that made possible to continue this project.
This study has been financially supported by ANTARES, the Italian
Ministry of University, Scientific Research and Technology (MURST), and the
Italian Space Agency (ASI). Finally, this study is part of the scientific
programme "Galaxy formation and evolution" financed by the European Community
with the TMR grant ERBFMRX-CT96-0086.


\begin{thebibliography}{}


\bibitem{} Arimoto, N.,  Yoshii, Y. 1987, A\&A,  173, 23

\bibitem{} Arimoto, N.,  Yoshii, Y. 1989, A\&A, 224, 361
  
\bibitem{} Balsara, D. S. 1995, J. Chem. Phys.,  121, 357

\bibitem{} Barnes, J., Hut, P, 1986, Nature,  324, 446

\bibitem{} Baugh C. M., Cole, S., Frenk, C.S. 1996, MNRAS, in press

\bibitem{} Becquaert J. F., Combes F. 1997, A\&A, in press

\bibitem{} Bender, R., 1997, in {\it The Nature of Elliptical Galaxies}, 
           Proceedings of the Second Stromlo Symposium, eds. M. Arnaboldi, 
           G.S. Da Costa \& P. Saha, in press  

\bibitem{} Beninc\`a, M., Carraro, G. 1995, in {\it Science and Supercomputing 
           at CINECA},   Report 1995, 27 

\bibitem{} Benz, W,, Bowers, R. L., Cameron, A. G. W., Press, W. H. 1989, 
           ApJ,  348, 650

\bibitem{} Benz, W, 1990, in {\it Numerical Modelling of Nonlinear Stellar 
           Pulsation}, ed. J. R. Buchler, p. 269,  Dordrecht: Kluwer 

\bibitem{} Bertelli, G., Bressan, A., Chiosi, C., Fagotto,  F., 
           Nasi, E, 1994, A\&AS, 106, 275

\bibitem{} Bonor, W. B. 1956, MNRAS,  116, 351

\bibitem{} Bressan, A., Chiosi, C., Fagotto, F. 1994, ApJS, 94, 63

\bibitem{} Bressan, A., Chiosi, C., Tantalo, R. 1996, A\&A, 311, 425
 
\bibitem{} Bruzual, G.,   Charlot S. 1993, ApJ, {\bf 405}, 538

\bibitem{} Caimmi, R., Secco, L. 1986, Astrophys. Space Sci. 119, 315

\bibitem{} Carraro, G., 1995, PhD. Thesis, Padova University 

\bibitem{} Carraro, G., Ng, Y.K., Portinari L., 1997, MNRAS submitted 
({\tt astro-ph/9707185})

\bibitem{} Chiosi, C., Bressan, A.,  Portinari, L.,  Tantalo, R.  1997,
           A\&A, submitted ({\tt astro-ph/9708123})

\bibitem{} Davis, M., Efstathiou, G., Frenk, C.S., White, S.D.M. 1992, 
           Nature,  356, 489
 
\bibitem{} Ebert, R, 1955, Zs. Ap., 37, 217

\bibitem{} Edvardsson B., Andersen J., Gustafsson B., et~al., 1993,
	A\&A 275, 101 

\bibitem{} Evrard, A. E, 1988, MNRAS.  235, 911

\bibitem{} Gibson, B.K. 1996a, ApJ, 468, 167

\bibitem{} Gibson, B.K. 1996b, ApJ, 468, 167

\bibitem{} Gibson, B.K. 1996c, MNRAS, 278, 829

\bibitem{} Gibson, B.K. 1997, preprint

\bibitem{} Gibson, B.K., Matteucci, F. 1997, ApJ, 475, 47

\bibitem{} Gingold, A. A., Monaghan, J. J. 1983, J. Comput. Phys.,
           46, 429

\bibitem{} Greggio, L, Renzini, A. 1983, A\&A, 118, 217

\bibitem{} Groom W. 1997, Phd Thesis

\bibitem{} Haehnelt, M., Steinmetz, M., Rauch, M. 1996a, ApJ, 465, L95

\bibitem{} Haehnelt, M., Steinmetz, M., Rauch, M. 1996b, MNRAS, in press 
 
\bibitem{} Hernquist, L, Katz, N. 1989, ApJS, 70, 419

\bibitem{} Hollenbach, D. 1988, Astro. Lett. and Communications, vol. 26, 191

\bibitem{} Hollenbach, D., McKee, C.F. 1979, ApJS, 41, 555

\bibitem{} Katz, N. 1992, ApJ, 391, 502

\bibitem{} Katz, N, Gunn, J. 1991, ApJ, 377, 365

\bibitem{} Kauffmann, G., Charlot, S., White, S.D.M.  1996, MNRAS, 283, L117

\bibitem{} Kauffmann, G., White, S.D.M.,  Guiderdoni, B.  1993, MNRAS, 264, 201

\bibitem{} Lia, C. 1996, Master Thesis, Padova University

\bibitem{} Matteucci, F. 1997, Fund. Cosmic Phys. in press 

\bibitem{} Miller, G. E., Scalo, FJ.M.. 1979, ApJS 41, 513 

\bibitem{} Monaghan, J. J., Lattanzio, J. C. 1985, A\&A, 149, 135

\bibitem{} Monaghan, J. J., Lattanzio, J. C. 1991, ApJ,  375, 177

\bibitem{} Navarro, J.F., Frenk, C.S., White, S.D.M. 1996, ApJ, 462, 563

\bibitem{} Navarro, J. F,, White, S. D. M. 1993, MNRAS, 265, 271

\bibitem{} Navarro, J. F., Steinmetz, M. 1996, Astro-ph 9605043

\bibitem{} Nelson, R. P.,  Papaloizou, J. C. B. 1994, MNRAS, 270, 1

\bibitem{} Olling, P. R. 1996, AJ, 112, 481

\bibitem{} Persic M., Salucci P., 1997 , in "Dark Matter in Galaxies",
proceeding of the Sesto conference.

\bibitem{} Portinari, L.,  Chiosi, C.,  Bressan, A. 1996, A\&A. submitted

\bibitem{} Raiteri C.M., Villata M., Navarro J.F., 1996, A\&A 315, 105


\bibitem{} Rampazzo, R., Longhetti, M., Bressan, A., Chiosi, C. 1997,
           preprint
           
\bibitem{} Sackett, P. D.,  Morrison, H. L.,  Harding, P., Boroson, T. A.
             1994, Nature, 370, 441

\bibitem{} Salpeter, E. E. 1955, ApJ, 121, 161

\bibitem{} Schmidt, M, 1959, ApJ, 129, 243

\bibitem{} Schweizer, F.,   Seitzer, P. 1992, AJ, 104, 1039 

\bibitem{} Steinmetz, M. 1996a, Proc. Int. School of Physics "Enrico Fermi" -
           {\it Dark Matter in the Universe}, Varenna, Italy, July 24 - 
           August 4 1995, IOP, Bristol

\bibitem{} Steinmetz, M. 1996b,  MNRAS, 278, 1005

\bibitem{} Steinmetz M. Bartelmann M., 1996, MNRAS 272, 570

\bibitem{} Steinmetz, M.,  M\"uller, E, 1993, A\&A.  268, 391

\bibitem{} Steinmetz, M.,  M\"uller, E, 1994, A\&A.  281, L97

\bibitem{} Steinmetz, M., M\"uller, E. 1995, MNRAS, 276, 459

\bibitem{} Sutherland, R. S., Dopita, M. A, 1993, ApJS, 88, 253

\bibitem{} Tantalo, R., Chiosi, C., Bressan, A., Fagotto, F. 1996, 
            A\&A, 311, 361

\bibitem{} Tantalo, R.,  Bressan, A.,  Chiosi, C.,  Portinari, L.
           1997, A\&A, to be submitted 

\bibitem{} Tegmark, M., Silk, J.,  Rees, M.J., Blanchard, A., Abel, T.,
            Palla F. 1996, ApJ, submitted

\bibitem{} Theis, Ch., Burkert, A., Hensler, G. 1992, A\&A, 265, 465

\bibitem{} Tinsley, B M, 1980, ApJ, 241, 41

\bibitem{} van den Bergh, S. McClure, R. D, 1994, ApJ, 425, 205

\bibitem{} White, S. D. M. 1984, ApJ, 286, 38

\bibitem{} Worthey, G. 1994, ApJS, 95, 107

\end{thebibliography}
\end{document}